\documentclass[twocolumn,superscriptaddress,amsmath,amssymb,aps,prx,floatfix]{revtex4-2}
\usepackage{graphicx}
\usepackage{color}
\usepackage{overpic}
\usepackage{braket}
\usepackage{bm}
\usepackage{comment}
\usepackage[breaklinks=true,colorlinks,citecolor=blue,linkcolor=blue,urlcolor=blue]{hyperref}
\usepackage{soul}

\usepackage{tabularx}  

\usepackage{booktabs}  
\usepackage{makecell}  

\begin{document}

\title{{Dicke superradiant heat current enhancement in circuit quantum electrodynamics}}

\author{Gian Marcello Andolina}
\affiliation{JEIP, UAR 3573 CNRS, Collège de France, PSL Research University, 11 Place Marcelin Berthelot,  F-75321 Paris, France}

\author{Paolo Andrea Erdman}
\affiliation{Freie Universit{\" a}t Berlin, Department of Mathematics and Computer Science, Arnimallee 6, 14195 Berlin, Germany}
\author{Frank No{\'e}}

\affiliation{Microsoft Research AI4Science, Karl-Liebknecht Str. 32, 10178 Berlin, Germany}
\affiliation{Freie Universit{\" a}t Berlin, Department of Mathematics and Computer Science, Arnimallee 6, 14195 Berlin, Germany}
\affiliation{Freie Universit{\" a}t Berlin, Department of Physics, Arnimallee 6, 14195 Berlin, Germany}

\affiliation{Rice University, Department of Chemistry, Houston, TX 77005, USA}
\author{Jukka Pekola}
\affiliation{Pico group, QTF Centre of Excellence, Department of Applied Physics,
Aalto University, P.O. Box 15100, FI-00076 Aalto, Finland}

\author{Marco Schir{\`o}}
\affiliation{JEIP, UAR 3573 CNRS, Collège de France, PSL Research University, 11 Place Marcelin Berthelot,  F-75321 Paris, France}

\begin{abstract}

Controlling the flow of energy and heat at the micro-scale is crucial to achieve energy-efficient quantum technologies, for on-chip thermal management, and to realize quantum heat engines and refrigerators.
Yet, the efficiency of current quantum technologies is affected by thermal noise, and efficient cooling of quantum devices remains challenging in various solid-state implementations such as superconducting circuits. 
Collective effects, such as the Dicke superradiant emission, have been exploited to enhance the performance of quantum devices. However, the inherently transient nature of Dicke superradiant emission raises questions about its impact on steady-state properties. 
Here, we study how to enhance the steady-state heat current flowing between a hot and a cold bath through an ensemble of $N$ qubits, that are collectively coupled to the thermal baths. Remarkably, we find a regime where the heat current scales quadratically with $N$ in a finite-size scenario.
Conversely, when approaching the thermodynamic limit, we prove that the collective scenario exhibits a parametric enhancement over the non-collective case.
We then consider the presence of a third uncontrolled {\it parasitic} bath, interacting locally with each qubit, that models unavoidable couplings to the external environment. Despite having a non-perturbative effect on the steady-state currents, we show that the collective enhancement is robust to such an addition.
Finally, we discuss the feasibility of realizing such a Dicke heat valve with superconducting circuits.     
Our findings indicate that in a minimal realistic experimental setting with two superconducting qubits, the collective advantage offers an enhancement of approximately $10\%$ compared to the non-collective scenario.

\end{abstract}

\maketitle

\section{Introduction}

Quantum thermodynamics \cite{Kosloff_Entropy_2013,benenti2017rep,Vinjanampathy_ContempPhys_2016,Goold_JPhysA_2016,QTD_book,Lostaglio_RPP_2019} is the study of heat and work management in quantum systems. Within the recent blooming of quantum technologies, the critical aspect of energy management becomes increasingly crucial\cite{Auffeves_PRXQuantum_2022}, due to the natural interest in building energy-efficient quantum technologies and limiting associated energy waste. In this context, it is particularly relevant to investigate quantum heat transport in superconducting circuit Quantum Electrodynamics (circuit QED)~\cite{Blais_RMP_2021} which is among the most promising platforms for quantum technologies and quantum computation~\cite{Arute_Nature_2019}.
The heat current flowing across circuit QED devices has been recently measured across various designs \cite{fornieri2017nattech, maillet2020natcom, ligato2022natphys, gubaydullin2022natcom,gumus2023}, and circuit QED is emerging as a platform to realize quantum heat engines and refrigerators \cite{Campisi_2013,karimi_2016,hofer_2016_prb,hofer_2016_prb2,sanchez_apl_2017,tan_2017,hardal_2017,funo_2019,tabatabaei_2022_prb,erdman_npj_2022,Guthrie_2022,aamir_arxiv_2023,manzano_prr_2023}.

Two recent circuit QED experiments have investigated the heat transport of microwave photons scattering off a single qubit, realizing a photonic heat valve~\cite{Ronzani_NatPhys_2018,senior2020commun}. In these experiments, a single transmon qubit was capacitively coupled to two microwave resonators, each of them in contact with a resistance acting as a heat bath. 
A natural question is therefore whether the performance of such a device could be enhanced in the presence of multiple qubits interacting among each other. Indeed, transport of heat and energy are well known to be sensitive probes of collective and many-body effects~\cite{pekola2021colloquium,fazio1998anomalous,kane1996thermal,jezouin2013quantum,bertini2020finitetemperature,larzul2023energy}. 
A notable example where collective phenomena result in an enhanced emission and {\it super-extensive} scaling is provided by the Dicke model
 \cite{Dicke54}, where an ensemble of $N$ atoms in an optical cavity collectively radiates with a superextensive intensity that scales as $N^2$, i.e. enhanced by a factor $N$ with respect to ordinary fluorescence, where atoms emit independently. In the Dicke model, the electrical dipoles of the atoms synchronize thanks to their collective coupling to the optical cavity modes, leading to an enhanced emission which has been dubbed ``superradiance''~\cite{Raimond_RMP_2001,Gross_PhysRep_1982}.
Superradiant emission has been observed in various systems, including Rydberg atoms in cavities \cite{Kaluzny_PRL_1983}, color centers in diamonds \cite{Angerer_NatPhys_2018}, and in superconducting qubits \cite{Wang_PRL_2020}.

Collective effects, including Dicke superradiance, have been proposed to improve the performance of thermometers \cite{correa2015prl,mehboudi2022prl,Abiuso_arXiv_2022}, of quantum heat engines \cite{PhysRevA.107.L040202,Campisi_NatComm_2016,Fusco_PhysRevE_2016,Niedenzu_NJP_2018,manzano_njp_2019,mitchison_2019,abiuso2020,souza_pre_2022,Liu_ApplPhysLett_2023,Kamimura_PRL_2023,Kamimura_PRL_2023,carrega_arXiv_2023,KenFunoPRL}, of quantum batteries \cite{Ferraro17,Andolina19,Quach,Campaioli_arXiv_2023,CampaioliPRL}, of refrigerators \cite{PhysRevApplied.19.034023}, of heat transfer\cite{VOGL},  of energy transfer between quantum devices \cite{CrescentePRR} and to reduce the dissipated work in finite-time thermodynamics \cite{rolandi2023prl}.

\begin{figure}[t!]
\vspace{2mm}
\centering
 \begin{overpic}[width=0.95\columnwidth]{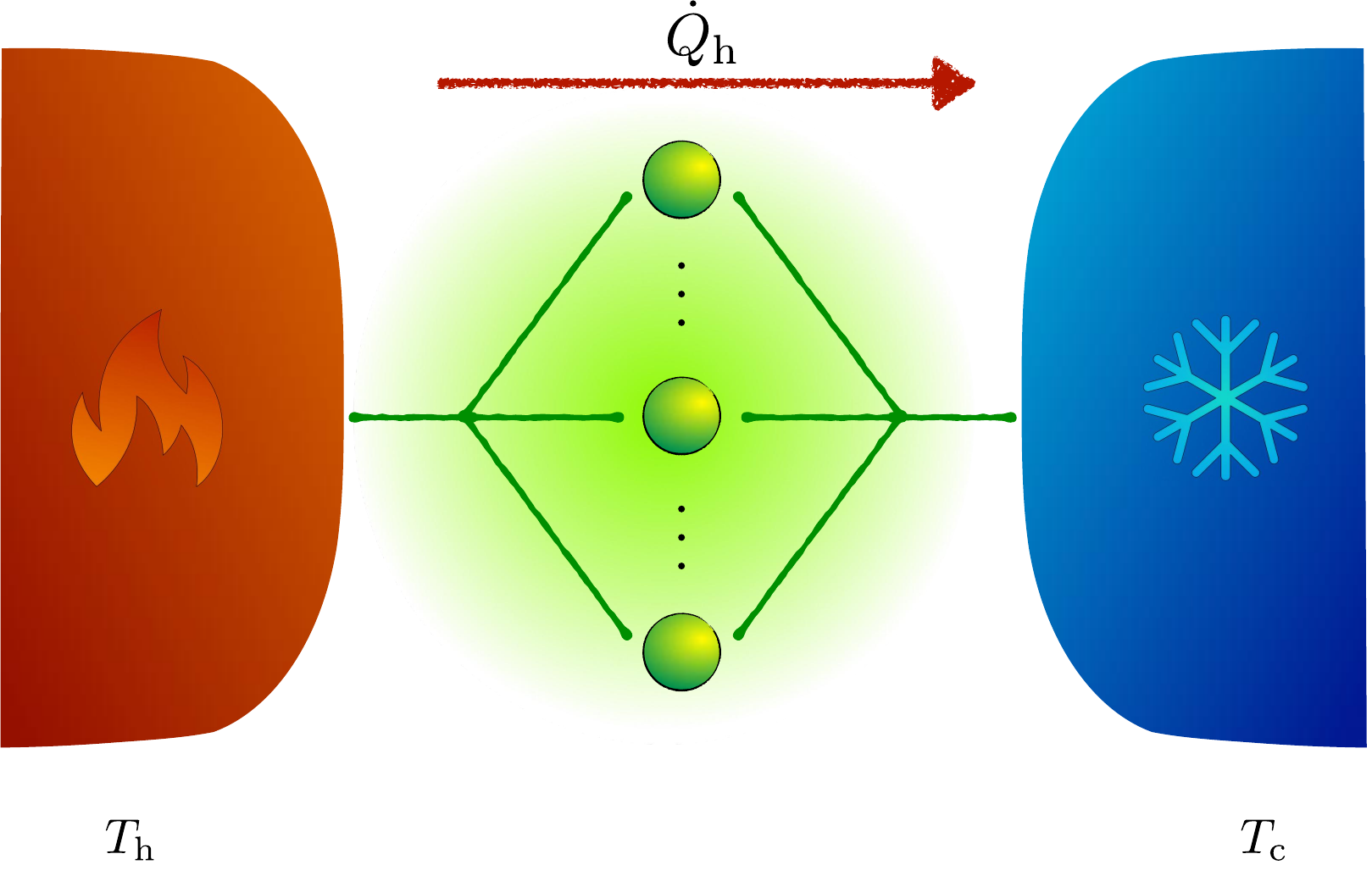}\put(-4,62){}
 \end{overpic}
\caption{(Color online) This figure shows an illustration of our system, where a heat current $\dot{Q}_{\rm h}$ flows from a hot bath, characterized by a temperature \(T_{\rm h}\), to a cold bath with a temperature \(T_{\rm c}\). This heat current is mediated by an ensemble of \(N\) qubits, that are not directly coupled. The collective coupling to the baths leads to a collective enhancement of the heat current.
}
\label{fig:sketch0}
\end{figure}

In this work, we investigate the role of collective superradiant coupling in enhancing the heat current flowing between two thermal baths through an ensemble of artificial atoms (qubits), as depicted in Fig.~\ref{fig:sketch0}. This is a non-trivial question, particularly since superradiance is typically a transient phenomenon observed in emission and absorption, whereas our focus is on a non-driven, steady-state situation. Remarkably, we find that even under these conditions, the collective coupling between the qubits and the thermal bath can lead to the emergence of super-extensive scaling in the heat current for systems of finite size. 
While the super-extensive scaling vanishes in the thermodynamic limit, we find that the heat current is nonetheless enhanced by a temperature-dependent prefactor that diverges as the temperature increases.

 While many previous proposals have examined the influence of collective effects on thermal machines \cite{correa2015prl,mehboudi2022prl,Abiuso_arXiv_2022,PhysRevA.107.L040202,Campisi_NatComm_2016,Fusco_PhysRevE_2016,Niedenzu_NJP_2018,manzano_njp_2019,mitchison_2019,abiuso2020,souza_pre_2022,Liu_ApplPhysLett_2023,Kamimura_PRL_2023,carrega_arXiv_2023,KenFunoPRL,Ferraro17,Andolina19,Quach,Campaioli_arXiv_2023,CampaioliPRL,PhysRevApplied.19.034023,CrescentePRR} and heat transport \cite{VOGL}, it remains crucial to discuss the robustness of these effects to noise. Noise can hinder the coherence of the dynamics, potentially undermining the collective enhancement.
Here, we show that this collective enhancement is robust to the addition of a third uncontrolled {\it parasitic} bath, which interacts locally with each qubit. This finding is crucial 
since, as we will show, the presence of an infinitesimally small local noise has a finite and non-perturbative effect on the steady-state heat currents, and removes the dependence of the steady-state heat current from the initial state preparation. Furthermore, we demonstrate the resilience of the superradiant effect in this realistic noisy environment.

Despite the intense theoretical interest in exploiting collective effects in quantum thermodynamical machine \cite{correa2015prl,mehboudi2022prl,Abiuso_arXiv_2022,PhysRevA.107.L040202,Campisi_NatComm_2016,Fusco_PhysRevE_2016,Niedenzu_NJP_2018,manzano_njp_2019,mitchison_2019,abiuso2020,souza_pre_2022,Liu_ApplPhysLett_2023,Kamimura_PRL_2023,Kamimura_PRL_2023,carrega_arXiv_2023,KenFunoPRL,Ferraro17,Andolina19,Campaioli_arXiv_2023,CampaioliPRL,PhysRevApplied.19.034023,CrescentePRR,VOGL}, only a few experiments have been conducted in this context \cite{Kim_NatPhoton_2022,Quach}. 
Hence, we discuss the feasibility of measuring such a Dicke enhancement of the heat current in an experimental setup.
Such a device can be realized within the framework of circuit-QED \cite{Ronzani_NatPhys_2018,Pekola_NatPhys_2015,Blais_RMP_2021,Krantz_ApplPhysRev_2019}, where \(N\) transmons are capacitively coupled to two $RLC$ circuits where the dissipative nature of a thermal bath stems from the presence of the resistive elements.
Our findings indicate that, using experimentally realistic parameters in the minimal case of $N=2$ superconducting qubits, the collectively enhanced heat current is approximately  $10\%$ higher than in scenarios where collective effects are absent. This highlights the potential of leveraging collective quantum behaviors in practical thermodynamic applications, offering a measurable enhancement over more traditional designs, and providing a platform to experimentally observe superradiant effects in measurable steady-state heat currents.
Our proposed device also represents a many-body collective version of a heat valve, a device that has been previously implemented in the context of circuit-QED  \cite{Ronzani_NatPhys_2018}. A heat valve is a device designed to control the flow of heat between two baths. Indeed, the circuit pictured in Fig.~\ref{fig:sketch} can function as a heat valve.  The ${\rm LC}$ circuits in Fig.~\ref{fig:sketch} (left and right elements) act as a filter, when the qubits are tuned to resonate at the $LC$ frequency, there is an efficient flow of heat between the two baths. Conversely, if the qubits are detuned from the $LC$ frequency, the heat flow is effectively impeded due to the frequency mismatch. This dynamic tuning capability allows for controlled manipulation of heat transfer, embodying the core function of a heat valve.


This article is organized as follows, in Sec.~\ref{sec:model} we derive the model and discuss the super-extensive behavior of the heat current in the noiseless case, i.e. when the modulus of the collective spin operator is conserved. In Sec.~\ref{sec:Parasitic_bath} we discuss the resilience of this super-extensive behavior to the addition of a realistic parasitic bath, breaking the conservation of the modulus of the collective spin. In Sec.~\ref{sect:EP} we discuss the experimental feasibility of observing our findings and we propose a minimal experimental setup that exhibits the super-extensive behavior of the heat current. In Sec.~\ref{sec:Conclusions} we draw our conclusions. Apps.~\ref{App:derivation_of_TC},\ref{App:derivation_Lindblad},\ref{appendixA}, \ref{appendixHeatDetuned} contain a series of technical details.

\section{The Dicke-superradiant heat valve}
\label{sec:model}

\begin{figure}[h!]
\vspace{2mm}
\centering
\begin{overpic}[width=0.95\columnwidth]{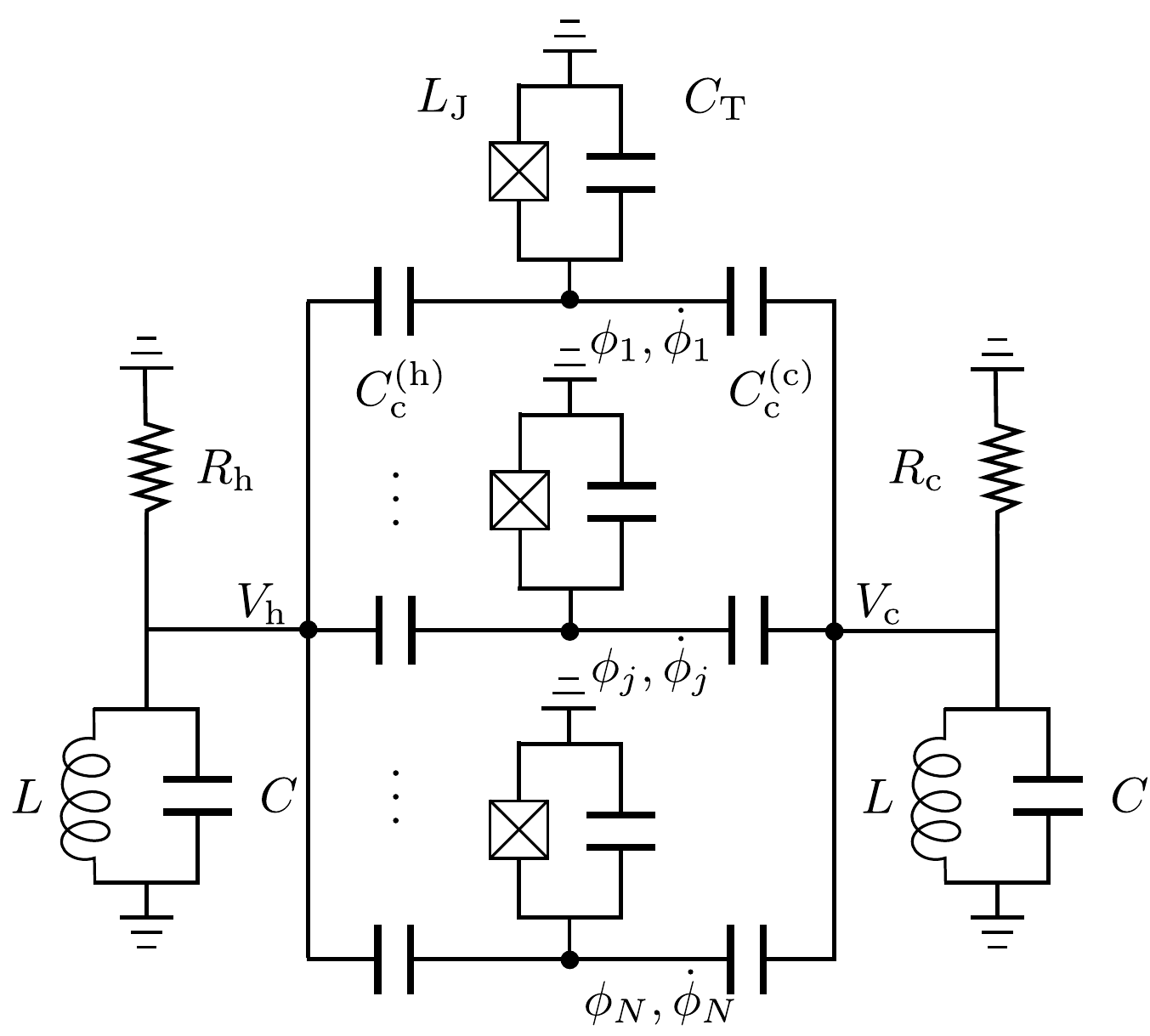}\put(2,60){}
 \end{overpic}
\caption{(Color online) This figure shows a lumped-element circuit diagram of a circuit-QED realization of this system. The circuit features two \(LC\) resonators, with associated voltages \( V_i\) where \(i={\rm h,c}\). Each resonator consists of an inductance \(L\) and a capacitance \(C\), and is coupled to a resistance \( R_i \). The ensemble of \(N\) transmons is represented by fluxes \(\phi_{j}\) for \(j=1,\ldots,N\), and their time derivatives \(\dot{\phi}_j\). Each transmon is made up of a Josephson junction with an associated inductance \(L_{\rm J}\) and a capacitance \(C_{\rm T}\). The transmons are capacitively coupled to the resonators via capacitors \(C_{\rm c}^{(i)}\).  }
\label{fig:sketch}
\end{figure}
 The lumped-element circuit diagram, including capacitances, inductances, resistances, and the various variables, is depicted in Fig.~\ref{fig:sketch}. 
 The quantum heat valve under study consists of \( N \) transmon qubits coupled to two resistors denoted by $R_i$ in parallel with \( LC \)-resonators. These \( RLC \) 
circuits form two heat baths characterized by their temperature $T_i$. We denote the two baths  as `hot' (h) and `cold' (c),  hence $i = \{\rm{h}, \rm{c}\}$ and $T_{h}\geq T_c$.
As derived in App.~\ref{App:derivation_of_TC}, the total Hamiltonian of the system can be written as
\begin{equation}
\label{eq:H_tot}
\hat{H}_{\rm tot} =\hat{H}_0+\hat{H}_{\rm int}+\hat{H}_{ RLC}\,.
\end{equation}
$\hat{H}_{0}$ and $\hat{H}_{\rm int}$ describe, respectively, the Hamiltonian of the qubits and their coupling to the $RLC$ circuit. Before projecting onto the qubit subspace, the Hamiltonian of the circuit is given by 
\begin{equation}
\begin{split}
\label{eq:H_circuit_main}
\hat{H}_{0} &=\frac{1}{2}\sum_{j=1}^N\left[\frac{1}{C_{\rm eff}} \hat{Q}_j^2- E_\text{J} \cos\left(\frac{2\pi\hat{\phi}_j}{\phi_0}\right)\right]~,
\\\hat{H}_{\rm int}&=-\sum_{j=1}^N\sum_{i={\rm h,c}}\frac{C^{(i)}_{\rm c}}{C_{\rm T}}\hat{Q}_j\hat{V}_i~, 
\end{split}
\end{equation}
where $C_{\rm eff}=C_{\rm T}\left[1-{\sum_{i={\rm h,c}}C^{(i)}_{\rm c}}/{C_{\rm T}}\right]^{-1}$, $\hat{Q}_j,\hat{\phi}_j$ are the charge and flux associated with the $j$-th transmon, $C_{\rm T}$ (\( E_\text{J} \)) is the transmon capacitance (Josephson energy), $\phi_0=h/(2e)$ , $C^{(i)}_{\rm c}$ are the coupling inductances, and $\hat{V}_i$ is the voltage of the $i$-th $RLC$ circuit (for $i=\text{h},\text{c}$), as reported in the lumped-circuit diagram in Fig.~\ref{fig:sketch}. By projecting the Hilbert space onto the two-level subspace of each transmon we obtain 
\begin{align}
\label{eq:H0}
\hat{H}_0=\sum_{j=1}^N \frac{\hbar\omega_0}{2}\hat{\sigma}^{(j)}_z,
\end{align}
where \( \hat{\sigma}^{(j)}_z \) represents the Pauli $z$-operator for the \( j \)-th qubit and  $\hbar \omega_{0}$ is the qubit energy, and
\begin{align}
\label{eq:Hint}
\hat{H}_{\rm int}=-\sum_{j=1}^N\sum_{i={\rm h,c}}{\hbar G_i} \frac{\hat{\sigma}^{(j)}_x}{2}\hat{V}_i,
\end{align}
where $\hat{\sigma}^{(j)}_x$ is the dipole operator of the \( j \)-th qubit (the Pauli $x$-operator) and \( G_i\) is the coupling strength that we consider different for the two $RLC$ circuits but otherwise uniform across the qubits.
This coupling plays a crucial role in the interaction dynamics of the qubits with the thermal baths.  Finally, $\hat{H}_{ RLC}$ describes the Hamiltonian of the two $RLC$ circuits as described in Appendix~\ref{App:derivation_of_TC}.

It is useful to introduce a collective spin operator $\hat{{J}}_{\alpha}$, with $\alpha=x,y,z$, as 
\begin{equation}
  \hat{{J}}_{\alpha} = \sum_{j=1}^N  \frac{\hat{{\sigma}}_{\alpha}^{(j)},}{2}~,
\end{equation}
where $\hat{{\sigma}}_{\alpha}^{(j)}$ are the Pauli operators for the $j$-th qubit.
The total Hamiltonian in Eq.~\eqref{eq:H_tot} can be thus re-written (up to an additive constant) as

\begin{equation}
\begin{split}
\label{eq:H_tot_J}
\hat{H}_{\rm tot} = \hbar \omega_0\hat{J}_z
-\sum_{i={\rm h,c}}{\hbar G_i} \hat{J}_x\hat{V}_i+\hat{H}_{ RLC}\,.
\end{split}
\end{equation} 
This Hamiltonian conserves the norm of the collective spin operator $\hat{J}^2$, where
\begin{equation}
  \hat{J}^2\equiv \sum_{\alpha=x,y,z}\hat{J}_\alpha^2~.
\end{equation}

It is thus useful to introduce the Dicke states, given by
\begin{equation}
\label{eq:dicke_states}
  |J, m_J\rangle = \sqrt{\frac{(J+m_J)!(J-m_J)!}{(2J)!}} \left(\hat{J}_+ \right)^{J+m_J} |J, -J\rangle~,
\end{equation}
where $J(J+1)$ ($m_J$) are the eigenvalues of $\hat{J}^2$ ($\hat{J}_z$), and $\hat{J}_+$ ($\hat{J}_-$) is the raising (lowering) operator.

Since the total Hamiltonian given by Eq.~\eqref{eq:H_tot_J} commutes with \( \hat{J}^2 \), the system dynamics is confined to a subspace characterized by a fixed value of \( J \). The allowed values for \( J \) are non-negative and follow the sequence \( J = N/2, (N-1)/2, (N-2)/2, \ldots \). Correspondingly, \( m_J \) can range from \( -J \) to \( J \) in integer steps, namely \( m_J = -J, -J+1, \ldots, J-1, J \).
As an example, consider a system of \( N=2 \) qubits. Here, the total collective spin \( J \) can either be 1 or 0. This means that the system can either be in a triplet state (\( J=1 \)) with \( m_J=-1, 0, 1 \) or in a singlet state (\( J=0 \)) with \( m_J=0 \).

Motivated by existing experimental setups \cite{Ronzani_NatPhys_2018}, we are interested in a scenario where the qubits are weakly coupled to the baths. In this regime, it is possible to invoke the Born-Markov approximation and describe the open system dynamics of the qubits with a suitable Lindblad master equation

\begin{equation}
\label{eq:Lindblad}
  \frac{d\hat{\rho}}{dt} = -\frac{i}{\hbar}[\hat{H}_0, \hat{\rho}] + \mathcal{D}(\hat{\rho})~,
\end{equation}
where $\hat{\rho}$ is the density operator of the qubits. A detailed derivation of this Lindblad master equation is carried out in Appendix \ref{App:derivation_Lindblad}, while here we summarize the results needed for our analysis. The dissipator $\mathcal{D}(\hat{\rho})$ arises from integrating out the $RLC$ circuits which are coupled to our system through the voltage $\hat{V}_i$ (see Eqs.~(\ref{eq:Hint}) and (\ref{eq:H_tot_J})). Within the Born-Markov approximation weak-coupling approximation, the dynamics of the voltage is fully encoded in the voltage dynamical structure factor \cite{Cattaneo_AdvQuantumTech_2021} $S_{\hat{V}_i,\hat{V}_j}(\omega)$, defined in Eq.~\eqref{eq: voltage dynamical structure factor}, which describe the voltage fluctuations occurring in the $RLC$ circuits.

The interaction term in Eq.~\eqref{eq:H_tot_J}, coupling the qubits system with the two resonators, is proportional to $\hat{J}_x=\hat{J}_++\hat{J}_-$.
Hence, after performing the secular approximation as needed to derive a Lindblad master equation \cite{Petruccione}, the suitable Lindblad operators for our system, describing transitions between the Dicke states induced by the exchange of energy with thermal baths, can be conveniently written using the collective spin raising and lowering operators $\hat{J}_+$ and $\hat{J}_-$. 
Hence, the total dissipator in Eq.~\eqref{eq:Lindblad} can be separated into contributions from the hot and cold baths:
\begin{equation}
\label{eq:total_dissipator}
  \mathcal{D}(\hat{\rho}) = \mathcal{D}_{\rm h}(\hat{\rho})+ \mathcal{D}_{\rm c}(\hat{\rho})~,
\end{equation}
 given by
\begin{equation}\label{eq:dissipator}
    \begin{split}
         \mathcal{D}_{i}(\hat{\rho})& = \gamma_{ i} (1 + n_{i}) \left( \hat{J}_- \hat{\rho} \hat{J}_+ - \frac{1}{2} \left\{ \hat{J}_+ \hat{J}_-, \hat{\rho} \right\} \right)+\\&+{\gamma_{\rm i} n_{i}}\left( \hat{J}_+ \hat{\rho} \hat{J}_- - \frac{1}{2} \left\{ \hat{J}_- \hat{J}_+, \hat{\rho} \right\} \right) ~,\\
         \vspace{0.3cm}
    \end{split}   
\end{equation}
 where $i={\rm h,c}$ and $\gamma_{\rm c}$ and $\gamma_{\rm h}$ are the transition rates for the cold and hot baths, respectively. 
The rates  \( \gamma_i \), that represent the coupling strength between the baths and the qubits, can be expressed in terms of the microscopic parameters of the circuit, as shown in Appendix \ref{App:derivation_Lindblad}. The thermal occupation numbers $n_i$ are given by the Bose–Einstein distribution
\begin{equation}
    n_{i}  = \frac{1}{\displaystyle \exp\left(\frac{ \hbar \omega_0}{k_{\rm B}T_i}\right) - 1}~,
    \vspace{0.2cm}
\end{equation}
where $k_{\rm B}$ is the Boltzmann's constant.

To summarize, after integrating out the two $RLC$ circuits representing the two thermal baths, we have obtained an effective description where the system of $N$ qubits is in contact with two structured thermal baths at finite temperature, each providing injection and depletion of energy through global spin-flip processes as described by Eq.~(\ref{eq:dissipator}). It is important to notice that the Lindblad master equation in Eq.~\eqref{eq:Lindblad} correctly inherits the conservation of $\hat{J}^2$ from the total Hamiltonian in Eq.~\eqref{eq:H_tot_J}. Specifically, this becomes a strong symmetry of the Lindbladian superoperator, since $\hat{J}^2$ commutes both with $H_0$ as well as with each jump operator $\hat{J}_{\pm}$. As such this corresponds to a conserved quantity of the Lindblad dynamics~\cite{albert2014symmetries}. 
Interstingly, analogous Lindblad master equations have been derived to describe dissipative (boundary) time crystals~\cite{iemini2018boundary}, where usually one considers a single bath at zero temperature. 

\subsection{Steady-state density matrix and effective temperature}

Since we are interested in computing the steady-state heat current flowing across the device, we start discussing the steady-state solution of the Lindblad master equation~(\ref{eq:Lindblad}). In the case of a single reservoir, i.e. $\mathcal{D}(\hat{\rho}) = \mathcal{D}_i(\hat{\rho})$ for $i=\text{h}$ or $i=\text{c}$, the steady-state solution corresponds to the thermal distribution (within each subspace with fixed $J$) characterized by the corresponding temperature $T_i$, since the rates satisfy the detailed balance condition
\begin{align}
  \frac{\gamma_i n_i}{\gamma_i (1 + n_i)} & = \exp\left(-\frac{ \hbar \omega_0}{k_{\rm B}T_i}\right)~,
\end{align}
capturing the equilibrium relation between absorption and emission rates in each bath. This result, consistent with the laws of thermodynamics, comes from using a master equation written in terms of the global jump operators for the coupled qubits (see Refs.~\onlinecite{hofer_2017_prb,cattaneo2019local,dechiara2018reconciliation} for a discussion of this point).
In the case of two baths, the total dissipator in Eq.~\eqref{eq:total_dissipator} can be expressed as:

\begin{widetext}
\begin{equation}
\label{eq:dissipator_global}
         \mathcal{D}(\hat{\rho}) =  \left[{\gamma_{\rm c} (1 + n_{\rm c})}+{\gamma_{\rm h} (1 + n_{\rm h})}\right]\left( \hat{J}_- \hat{\rho} \hat{J}_+ - \frac{1}{2} \left\{ \hat{J}_+ \hat{J}_-, \hat{\rho} \right\} \right)  + \left[{\gamma_{\rm c} n_{\rm c}}+{\gamma_{\rm h}  n_{\rm h}}\right]\left( \hat{J}_+ \hat{\rho} \hat{J}_- - \frac{1}{2} \left\{ \hat{J}_- \hat{J}_+, \hat{\rho} \right\} \right)~.
\end{equation}
\end{widetext}
For clarity in this discussion, we introduce a temperature \( T_0 \) associated with the qubit frequency $\omega_0$, \( T_0 \equiv (\hbar\omega_0)/k_{\rm B} \).
The dissipator in Eq.~(\ref{eq:dissipator_global}) defines a detailed balance equation

\begin{align}
\label{eq:betastar_implicit}
  \frac{{\gamma_{\rm c} n_{\rm c}}+{\gamma_{\rm h}  n_{\rm h}}}{{\gamma_{\rm c} (1 + n_{\rm c})}+{\gamma_{\rm h} (1 + n_{\rm h})}} & = \exp\left(-\frac{ T_0}{T^*}\right)~
\end{align}
which, in turn, defines an effective temperature $T^*$, such that $ T_c \leq T^* \leq T_h$. We can invert the previous equation to get an explicit expression for the effective temperature

%


\begin{equation}
\label{eq:betastar}
{T^*} = \frac{\displaystyle T_0}{
    \displaystyle \log \left[
        \frac{
            \gamma_{\text{c}} (1 + n_{\text{c}}) + \gamma_{\text{h}} (1 + n_{\text{h}})
        }{
            \gamma_{\text{c}} n_{\text{c}} + \gamma_{\text{h}} n_{\text{h}}
        }
    \right]
}~.
\end{equation}
%

The steady-state solution $\hat{\rho}^{(\rm s)}$, determined setting $\mathcal{D}(\hat{\rho}^{(\rm s)})=0$, is thus a thermal state at temperature $T^*$ within each subspace with fixed $J$.
However, since $\hat{J}^2$ is conserved, the relative occupation of different subspaces is fixed by the initial state.
Assuming that the initial state does not contain any coherence between subspaces with different $J$, the stationary state $\hat{\rho}^{(\rm s)}$ is given by

\begin{align}
\label{eq:rho}
\hat{\rho}^{(\rm s)}=\sum_{J} P_J \sum_{m_J} P(m_J|J)\ket{J,m_J}\bra{J,m_J} ~,
\end{align}
where $m_J$ are the eigenvalues of $\hat{J}_z$ compatible with $J$, $P_J$ is the occupation probability of subspace $J$ determined by the initial state, and $\ket{J,m_J}$ are the Dicke states, given in Eq.~\eqref{eq:dicke_states}. 
 It should be noted that the summation over \(J\) in Eq.~\eqref{eq:rho} incorporates the degeneracies associated with \(J\), as customary in angular momentum composition \cite{Sakurai1994}. For example, considering a system composed of three qubits, each characterized by a spin of \(\frac{1}{2}\), the resultant collective angular momentum can exhibit values of $J=1/2,1/2,3/2$, with the $J=1/2$ appearing twice. This is due to the composition rule that three spin-\(\frac{1}{2}\) entities combine to yield collective angular momenta of \(\frac{3}{2}\) and two instances of \(\frac{1}{2}\). This composition rule is customarily represented as $\frac{1}{2} \otimes \frac{1}{2} \otimes \frac{1}{2} = \frac{3}{2} \oplus \frac{1}{2} \oplus \frac{1}{2}$.
The dissipator in Eq.~\eqref{eq:dissipator_global} fixes the occupation within each subspace $J$, i.e. it fixes the conditional probabilities $P(m_J|J)$ to be thermal at the effective temperature $T^*$:

\begin{equation}
\label{eq:thermal_probabilities}
P(m_J \,|\, J) = \frac{ \displaystyle \exp \left( -\frac{m_J T_0}{ T^*} \right) }{Z_{J, T^*}}~,
\end{equation}
where $Z_{J,T^*}=\sum_{m_J} \exp\left[{-m_J T_0}/T^*\right]$ is the partition function.

\subsection{Heat Current}

The general expression for the total heat current $\dot{Q}$ flowing out of the thermal baths  can be defined as \cite{alicki1979,QTD_book}

\begin{align}
\label{eq:heatcurrent_def}
  \dot{Q}  \equiv \text{Tr}\left[ \hat{H}_0 \frac{d\hat{\rho}}{dt}\right] = \text{Tr}\left[ \hat{H}_0 \mathcal{D}(\hat{\rho}) \right]~.
\end{align}
The second equality has been obtained by enforcing the dynamics of the Lindblad master equation, Eq.~\eqref{eq:Lindblad}.
The total heat current $\dot{Q}$ can be split into two contributions corresponding to the different baths
\begin{align}
\label{eq:heatcurrent1}
 \dot{Q} = \dot{Q}_{\rm h}+ \dot{Q}_{\rm c}~,
\end{align}
where 
\begin{align}
\label{eq:heatcurrent2}
\dot{Q}_{i}\equiv \text{Tr}\left[ \hat{H}_0 \mathcal{D}_i(\hat{\rho}) \right]~.
\end{align}

Here, we are interested only in the steady state dynamics where $ \dot{Q}=0$ and $\dot{Q}_{\rm h}=- \dot{Q}_{\rm c}$.

Using the specific forms of the dissipators for the cold and hot baths (in Eq.~\eqref{eq:dissipator}), as detailed in App.~\ref{appendixA}, we arrive at the following expressions for the heat currents

\begin{align}
\label{eq:heatcurrent3}
 \dot{Q}_{i} & = \hbar \omega_0\gamma_{i} \left[- (1 + n_{i}) \langle  \hat{J}_+ \hat{J}_- \rangle +  n_{i} \langle  \hat{J}_- \hat{J}_+ \rangle \right]~, 
\end{align}
where $\langle \hat{x} \rangle \equiv \text{Tr}[\hat{\rho} \hat{x}]$. This result is reminiscent of the well-known superradiant energy emission 
in an ensemble of excited qubits~\cite{Raimond_RMP_2001,Gross_PhysRep_1982,Yan23,Masson_arXiv_2023}. In such a context, the system is typically considered to be in contact with a single cold bath at zero temperature, corresponding to setting $\gamma_{\rm h}=0$ in our case, and is initially prepared in a highly excited state, such as the state with $J=N/2$ and $M=J$. Consequently, the superradiant heat current flowing into the cold bath is given by $\dot{Q}_{\rm c}=-\hbar \omega_0\gamma_{\rm c}  \langle  \hat{J}_+ \hat{J}_- \rangle$. The term $\langle \hat{J}_+ \hat{J}_- \rangle$ can be expressed in terms of the individual qubit raising and lowering operators, $\hat{\sigma}_+^{(j)}$ and $\hat{\sigma}_-^{(j)}$, as $\sum_{j,j'} \langle \hat{\sigma}^{(j)}_+ \hat{\sigma}^{(j')}_- \rangle$.
This summation encompasses $N$ local population terms ($j=j^\prime$), and $\sim N^2$ non-local coherent terms ($j \neq j^\prime$) that may enable a {\it super-extensive} scaling effect.
However, it is important to notice that this superradiant effect is only present in the transient dynamics. Over time, the qubits will eventually reach thermal equilibrium with the zero-temperature bath, resulting in the cessation of the heat current.

Going back to the case under study, our primary focus is on the steady-state heat current that flows between the two baths.
 Using that $\dot{Q}_\text{c} = -\dot{Q}_\text{h}$, we can express the heat current as

\begin{align}
\label{eq:Q_steady}
    &\dot{Q}_\text{h} = \left({\frac{1}{\gamma_\text{h}} + \frac{1}{\gamma_\text{c}}}\right)^{-1}{\left(\frac{1}{\gamma_\text{h}} \dot{Q}_\text{h} - \frac{1}{\gamma_\text{c}} \dot{Q}_\text{c}\right)}=  \\
   &=\frac{2\gamma_\text{h}\gamma_\text{c}}{\gamma_\text{h}+\gamma_\text{c}}\hbar\omega_0  \langle -\hat{J}_z\rangle \left( n_\text{h} - n_\text{c}\right)~, \label{eq:heatcurrent4}
\end{align}
where we used that $[\hat{J}_+,\hat{J}_-] = 2\hat{J}_z$. Notice that the case of a single qubit can be obtained from this equation 
 setting $J=1/2$.

Notably, Eq.~\eqref{eq:heatcurrent4} reveals that the steady-state heat current does not explicitly depend on the sum of the non-local coherent terms that are present in $\langle \hat{J}_+ \hat{J}_- \rangle$ and $\langle \hat{J}_- \hat{J}_+ \rangle$, which are typically associated with a super-extensive behavior in the context of superradiant emission.
 Instead, the steady-state heat current is controlled by the operator $\hat{J}_z$, which can be represented as a sum of $N$ local terms, $\hat{J}_z=\sum_{j=1}^N \hat{\sigma}^{(j)}_z$. This would suggest the absence of a super-extensive scaling due to superradiant effects in this context.
Indeed, upper bounding the expectation value $\langle -\hat{J}_z\rangle$ in Eq.~(\ref{eq:heatcurrent4}) with $N/2$, we find the following exact upper bound 
\begin{equation}
    \dot{Q}_\text{h} \leq \dot{Q}_h^{\text{max}} \equiv 
    \frac{\gamma_\text{h}\gamma_\text{c}}{\gamma_\text{h}+\gamma_\text{c}} \hbar\omega_0 N (n_\text{h}-n_\text{c})
    \label{eq:q_max}
\end{equation}
revealing that, for large values of $N$, the heat current cannot scale more than linearly in the number of qubit. It is interesting to notice that the bound in Eq.~(\ref{eq:q_max}), which is saturable in the thermodynamic limit $N\to\infty$ (see Subsec.~\ref{ss:fixedsub}), has a ``Landauer-B{\"u}ttiker-like'' dependence on the temperature of the baths \cite{buttiker_prb_1985,landauer_1957}, and is analogous to the heat current flowing through a harmonic oscillator coupled to bosonic heat baths \cite{segal_prl_2005} - but enhanced by a factor $N$. Intuitively this can be expected because, in the thermodynamic limit, the collective spin operators effectively become bosonic operators \cite{ressayre_1975}, and should thus behave similarly to a harmonic oscillator. However, the rates are here increased by a factor $N$ since the qubits are coupled in parallel.

Surprisingly, despite the linear scaling of $\dot{Q}_\text{h}^{\text{max}}$ with $N$, below we show that the heat current $\dot{Q}_\text{h}$ can nonetheless exhibit a super-extensive scaling in the finite-size regime.
Indeed, if the qubit were in a global thermal state, it can be shown from Eq.~(\ref{eq:heatcurrent4}) that the scaling would be linear, since the thermal state is a tensor product of local qubit states. However, since $\hat{J}^2$ is conserved, the steady-state density matrix $\hat{\rho}^{\rm( s)}$, as indicated by Eq.~\eqref{eq:rho}, may not be purely thermal, and the presence of coherence between different qubits can enable a non-linear dependence of the heat current on the qubit number $N$.

As a benchmark for our study, we will compare our results to the case where \( N \) qubits (with equal frequency $\omega_0$) are independently coupled to a hot and cold bath. This scenario serves as a reference case, allowing for a clear comparison of our collective case described by a collective Lindblad master equation, Eq.~\eqref{eq:Lindblad}. To compute the heat current associated with a single qubit, we employ the Lindblad master equation in Eq.~\eqref{eq:Lindblad}, along with the expression of the steady-state heat current given in Eq.~\eqref{eq:heatcurrent4}, by setting $\bar{J}=1/2$, corresponding to $N=1$.
As all qubits in our model are identical, the total heat current in this independent case, denoted as $\dot{Q}^{\rm ind}_{\rm h}$, is determined by multiplying the heat current of a single qubit by the total number of qubits, $N$. This leads us to the following expression for the total heat current in the independent scenario:
\begin{align}
\label{eq:current_independent}
\dot{Q}^{\rm ind}_{\rm h} = N\hbar \omega_0 \left[ \frac{\gamma_{\rm h} \gamma_{\rm c}}{\gamma_{\rm c}(2n_{\rm c}+1)+\gamma_{\rm h}(2n_{\rm h}+1)}\right](n_{\rm h}-n_{\rm c})~.
\end{align}
An analogous results for the $N=1$ was previously derived in the spin-boson model \cite{segal_prl_2005,bhandari_prb_2021}.

\subsection{Results for the heat current in a fixed subspace}
\label{ss:fixedsub}

For the remainder of this section, we focus on scenarios where the initial state has a well-defined \( J = \bar{J} \), meaning that 

\begin{equation}
    \label{eq:Pdelta}
     P_J = \delta_{J,\bar{J}}~,
\end{equation}
in Eq.~\eqref{eq:rho}. Notice that, in general, $\bar{J}\leq N/2$, and the case $\bar{J}=N/2$ can be implemented by initializing all the qubits in their ground states \( \bigotimes_{j=1}^N \ket{0}_j \),  corresponding to the Dicke state \( \ket{N/2, -N/2} \) and hence the system's dynamics is constrained to the subspace with \( J = N/2 \). 
While this simplified case may not be directly applicable in experimental settings, it is a useful starting point for gaining insights into this problem before exploring more realistic scenarios.

In this setting, the steady-state heat current $\dot{Q}_{\rm h}$, as given in Eq.~\eqref{eq:heatcurrent4}, can be analytically derived by averaging the current operator over the thermal state at an effective temperature denoted by \( T^* \) and given in Eq.~(\ref{eq:betastar}). In the general case, the detailed calculation of this quantity is provided in Appendix~\ref{appendixA}. 

Analogously to the derivation of Eq.~(\ref{eq:q_max}), we can upper bound the expectation value $\langle -\hat{J}_z\rangle$ in Eq.~(\ref{eq:heatcurrent4}) with the conserved value of $\bar{J}$, yielding a generally tighter bound
\begin{equation}
    \dot{Q}_\text{h} \leq \dot{Q}_\text{h}^{\text{lim}} \equiv 
    \frac{2 \gamma_\text{h}\gamma_\text{c}}{\gamma_\text{h}+\gamma_\text{c}} \hbar\omega_0 \bar{J} (n_\text{h}-n_\text{c}).
    \label{eq:Q_steady_limit1}
\end{equation}
Interestingly, as derived in Eq.~(\ref{eq:limit1}), the heat current saturates this bound in the limit $\bar{J} T_0  \gg T^*$.  Physically, this can be interpreted as the low-temperature regime, i.e. when the effective temperature $T^*$ is considerably lower than $\bar{J} T_0$, or as the thermodynamic limit $N\to\infty$, choosing $\bar{J}=N/2$. Indeed, in these cases, the majority of spins are in the inverted state, corresponding to $\langle\hat{J}_z\rangle\approx -\bar{J}$. Since $\bar{J}$ is at most linear in $N$, this reveals that the heat current cannot scale super-extensively in the number of qubits both in the low-temperature regime and in the thermodynamic limit.

Nonetheless, in the thermodynamic limit $N\to \infty$ we still find a parametric advantage in the heat current. In this limit, the ratio between the heat current in the collective case $\dot{Q}_\text{h}^{\text{lim}}$ in the $\bar{J}=N/2$ case, and the heat current in the independent case $\dot{Q}^{\rm ind}_\text{h}$, is given by Eq.~(\ref{eq:limit3}), i.e.
\begin{equation}
\label{eq:Q_steady_limit3}
\frac{\dot{Q}_\text{h}^{\text{lim}}}{\dot{Q}^{\rm ind}_\text{h}}= \coth\left[\frac{T_0}{2T^*}\right]~.
\end{equation}
The ratio ${\dot{Q}^{\rm lim}_\text{h}}/{\dot{Q}^{\rm ind}_\text{h}}$ is always greater than $1$, indicating that the collective scenario yields a strictly larger heat current than the independent case. In the limit $T_0/T^*\gg 1$, the ratio approaches one. Conversely, in the limit where $T_0/T^*\ll 1$, the ratio becomes ${\dot{Q}^{\rm lim}_\text{h}}/{\dot{Q}^{\rm ind}_\text{h}}\approx \left(2T^*/T_0\right)$, which can be arbitrarily large. This indicates that, even in the thermodynamic limit, collective effects can provide an unbounded parametric enhancement to the heat current.

We now consider the high-temperature regime, i.e. $\bar{J} T_0  \ll T^*$. Using Eq.~\eqref{eq:limit2}, we find that
\begin{equation}
\label{eq:Q_steady_limit2}
\dot{Q}_\text{h}\approx \frac{2}{3}\frac{\gamma_\text{h}\gamma_\text{c} }{\gamma_\text{h}+\gamma_\text{c}} \frac{T_0}{T^*} (\hbar\omega_0) \bar{J} (1 + \bar{J})\left( n_\text{h} - n_\text{c}\right)~.
\end{equation}
In this scenario, the heat current exhibits a super-linear scaling with $ \bar{J}$. This scaling behavior is derived from the non-linear dependence of the thermal populations $P(m_{\bar{J}} \,|\, \bar{J})$ on $\bar{J}$, given in Eq. \eqref{eq:thermal_probabilities}, as shown in  Appendix~\ref{appendixA}.  In the case $\bar{J}=N/2$, this corresponds to a super-extensive scaling in the number of qubits $N$.
However, it is crucial to recognize that in the thermodynamic limit ($N\to\infty$), the condition $\bar{J} T_0  \ll T^*$ is never met when $\bar{J}=N/2$ since, as $N$ approaches infinity, $\bar{J}=N/2$ eventually surpasses $T^*/T_0$, and the upper limit $Q_\text{h}^{\text{lim}}$ in Eq.~(\ref{eq:Q_steady_limit1}), linear in $\bar{J}$, becomes tight.

\begin{figure}[t!]
\centering
  \begin{overpic} [width=0.95\columnwidth]{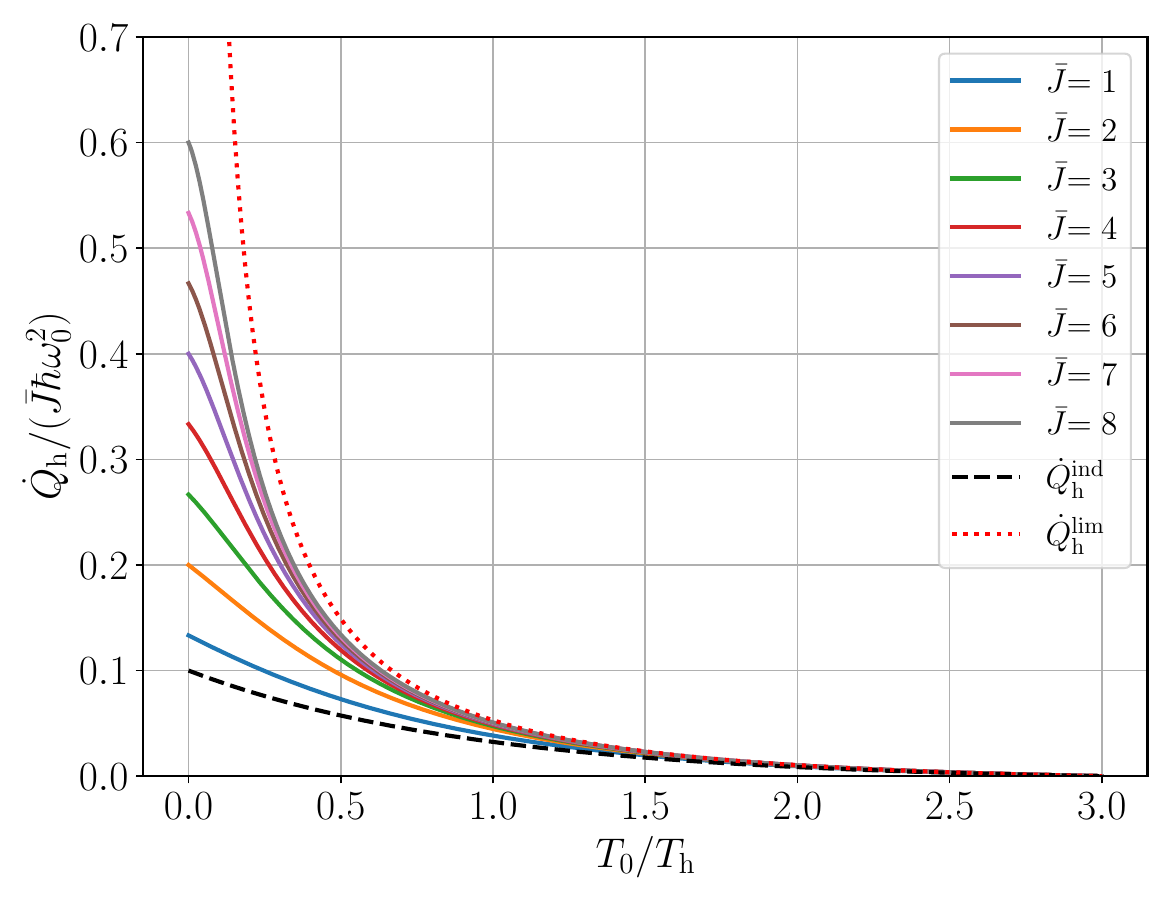}
    \put(0,73){(a)}
  \end{overpic}
   \begin{overpic}[width=0.95\columnwidth]{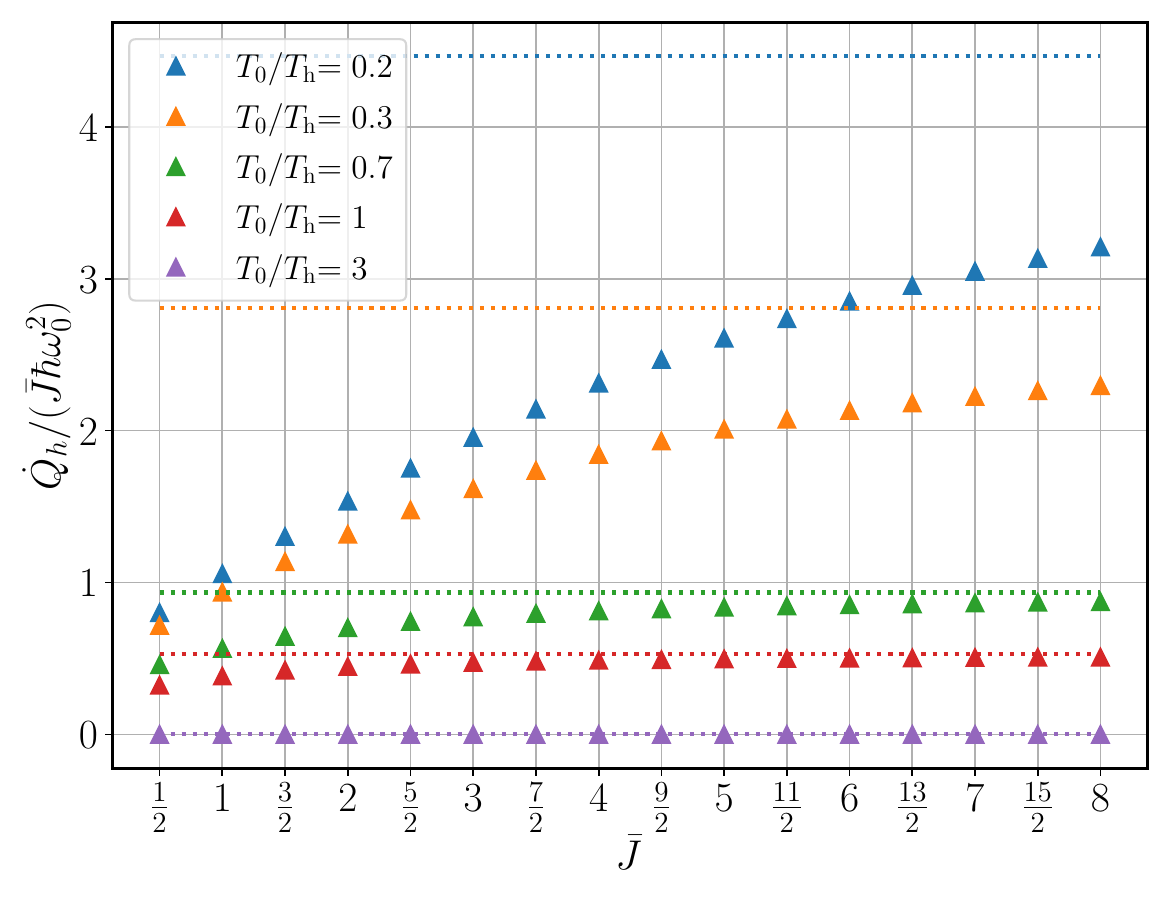}
    \put(0,73){(b)}
  \end{overpic}


\caption{ (Color online) Panels (a) shows the heat current as a function of the ratio $T_0/T_{\rm h}$ (where $T_0\equiv \hbar\omega_0/k_{\rm B}$) for various values of $\bar{J}$. 
The black dashed line indicates the independent current $\dot{Q}_{\rm h}^{\rm ind}$, whereas the red dotted curve represents the heat current in the thermodynamic limit, denoted as $\dot{Q}_{\rm h}^{\rm lim}$.
In panel (b), the heat current is plotted as a function of $\bar{J}$ for various ratios of $T_0/T_{\rm h}$. The dotted lines in this panel correspond to the heat current in the thermodynamic limit, $\dot{Q}_{\rm h}^{\rm lim}$, in the case $\bar{J}=N/2$. Each of these dotted lines corresponds to the same temperature ratio $T_0/T_{\rm h}$ of the data points of matching color.
The temperature of the cold bath is chosen to be $T_0/T_{\rm c}=3$ in both panels. Other parameters are $\gamma_{\rm h}=\gamma_{\rm c}=\omega_0/10$. 
}
\label{fig1}
\end{figure}

In Fig.~\ref{fig1} (a,b) we plot the steady-state heat current $\dot{Q}_{\rm h}$ (normalized by $\bar{J}\hbar\omega_0^2$) \eqref{eq:dissipator_global} focusing on system sizes of experimental relevance, specifically considering a system size up to $N=16$, which corresponds to $\bar{J}=8$. 
For convenience, we fix the temperature of the cold bath to be $T_c=T_0/3$ and tune the ratio $T_0/T_{\rm h}$, which according to the discussion on the steady-state's effective temperature allows tuning the effective temperature across the relevant scale $T_0$ as discussed above. In panel (a) we see that in the low-temperature regime, $T_0>1.5T_{\rm h}$ (corresponding to $T_0>T^*$ ), the ratio $\dot{Q}_{\rm h}/(\bar{J}\hbar\omega_0^2)$ does not depend on $\bar{J}$, i.e. the heat current scales linearly in $\bar{J}$, thus in $N$. However, in the high-temperature regime, i.e. $T_0 < 1.5T_{\rm h}$ (corresponding to $T_0 < T^*$), a super-linear scaling emerges. In panel (a) we also show as a dotted line the heat current obtained in the thermodynamic limit $\dot{Q}_{\rm h}^{\rm lim}$ (taking $\bar{J}=N/2$), as given in Eq.~\eqref{eq:Q_steady_limit1}, which is also an upper bound to the heat current. While, as expected, a saturation to the extensive limit $\dot{Q}_{\rm h}^{\rm lim}$ is seen for low-temperatures, there is no saturation for very high-temperatures, $T_0/T_h \ll 1$, where the superlinear scaling persists up to $N=16$. The scaling with the size is explicitly shown in panel (b), where the dependence of the ratio $\dot{Q}_{\rm h}/(\bar{J}\hbar\omega_0^2)$ on $\bar{J}$ is plotted for different temperatures, highlighting the superlinear scaling for small values of $T_0/T_{\rm h}$. Again, the heat current in the thermodynamic limit $\dot{Q}_{\rm h}^{\rm lim}$ is shown as dotted lines of correspondent color. Analogously to panel (a), for low-temperatures $T_0/T_{\rm h}\geq0.7$, the heat current saturates the bound given by $\dot{Q}_{\rm h}^{\rm lim}$. Instead, in the high-temperature regime, i.e. for $T_0/T_{\rm h}\leq 0.3$ and for realistic system sizes, the heat current always shows a superlinear behavior, far from saturation.

To summarize, for experimentally relevant cases where \(N \leq 10\), a super-extensive scaling of the heat current is always achievable provided the hot bath is sufficiently warm (high-temperature regime). In these scenarios, the value of \(\dot{Q}_{\rm h}^{\rm lim}\) essentially serves as an upper limit. 
Below we show that such a temperature regime that yields a super-extensive scaling of the heat current is within experimental reach and importantly, this behavior remains robust in the presence of noise.

\section{Impact of a parasitic bath}
\label{sec:Parasitic_bath}
The dynamics governed by the Lindblad master equation in Eq.~\eqref{eq:Lindblad} preserves the norm of the collective spin operator, \( \hat{J}^2 \). Consequently, the steady-state current depends on the probabilities \( P_J \), which are fixed by the initial state as described in Eq.~\eqref{eq:rho}. However, a dependence of the steady-state heat current on the initial state preparation of the qubits is not realistic in practice. Although short-term dynamics could be influenced by the initial state preparation, the long-term steady state is likely to be dominated by factors such as noise and local dissipation. These effects would eventually erase the memory of the initial state. Indeed, as we now show, such {\it parasitic} effects have a large impact on the steady-state heat current, even in the limit of vanishing small parasitic coupling, thus a non-perturbative effect.
To address this effect, we introduce a third, {\it parasitic} thermal bath that interacts {\it locally} with each qubit. Unlike the collective interactions from the primary thermal baths, this parasitic bath interacts uniformly but locally with each qubit, thereby breaking the conservation of \( \hat{J}^2 \). This parasitic bath is characterized by a temperature \( T_{\rm p} \) and a coupling strength \( \gamma_{\rm p} \).

The dissipator associated with the parasitic bath, \(\mathcal{D}_{\rm p}(\hat{\rho})\), can be expressed as

\begin{equation}
\mathcal{D}_{\rm p}(\hat{\rho}) =\sum_{j=1}^N\mathcal{D}^{(j)}_{\rm p}(\hat{\rho})~,
\end{equation}
and 
\begin{align}
\label{eq:Dissipator_independent_parasitic}
  & \mathcal{D}^{(j)}_{\rm p}(\hat{\rho}) = \gamma_{\rm p} n_{\rm p} \left( \hat{\sigma}_+^{(j)} \rho \hat{\sigma}_-^{(j)} - \frac{1}{2} \{ \hat{\sigma}_-^{(j)} \hat{\sigma}_+^{(j)}, \rho \} \right)+  \nonumber\\
  & +\gamma_{\rm p} (1 + n_{\rm p}) \left( \hat{\sigma}_-^{(j)} \rho \hat{\sigma}_+^{(j)} - \frac{1}{2} \{ \hat{\sigma}_+^{(j)} \hat{\sigma}_-^{(j)}, \rho \} \right) ~.
\end{align}
Here, $ n_{\rm p} = {1}/\{\exp\left[\hbar\omega_0/(k_{\rm B}T_{\rm p})\right] - 1\}$ represents the mean thermal occupation number of the parasitic bath.
The overall dissipative dynamic of the system influenced by all three baths is thus described by

\begin{equation}
\label{eq:DissipatorTot}
  \frac{d\hat{\rho}}{dt} = -\frac{i}{\hbar}[\hat{H}_0, \hat{\rho}] + \mathcal{D}_{\rm h}(\hat{\rho}) + \mathcal{D}_{\rm c}(\hat{\rho}) + \mathcal{D}_{\rm p}(\hat{\rho})~.
\end{equation}
 Although we assume that this parasitic bath is weakly coupled to the system compared to the primary baths (\( \gamma_{\rm p} \ll \gamma_{\rm h}, \gamma_{\rm c} \)), its main effect is the breaking the conservation of \( \hat{J}^2 \).
 Therefore, while the coupling strength \(\gamma_{\rm p}\) can be neglected when calculating the population within a given subspace, its influence is non-perturbative in determining the steady-state probabilities \( P_J \) of different subspaces. Hence, once we extract the values of $P_J$ from the numerics, we can still use the results of Sect.~\ref{ss:fixedsub} to determine the occupations $P(m_J|J)$ inside a given subspace.

\begin{figure*}[th]
\centering

\begin{minipage}{.49\textwidth}
  \centering
  \begin{overpic}[width=\linewidth]{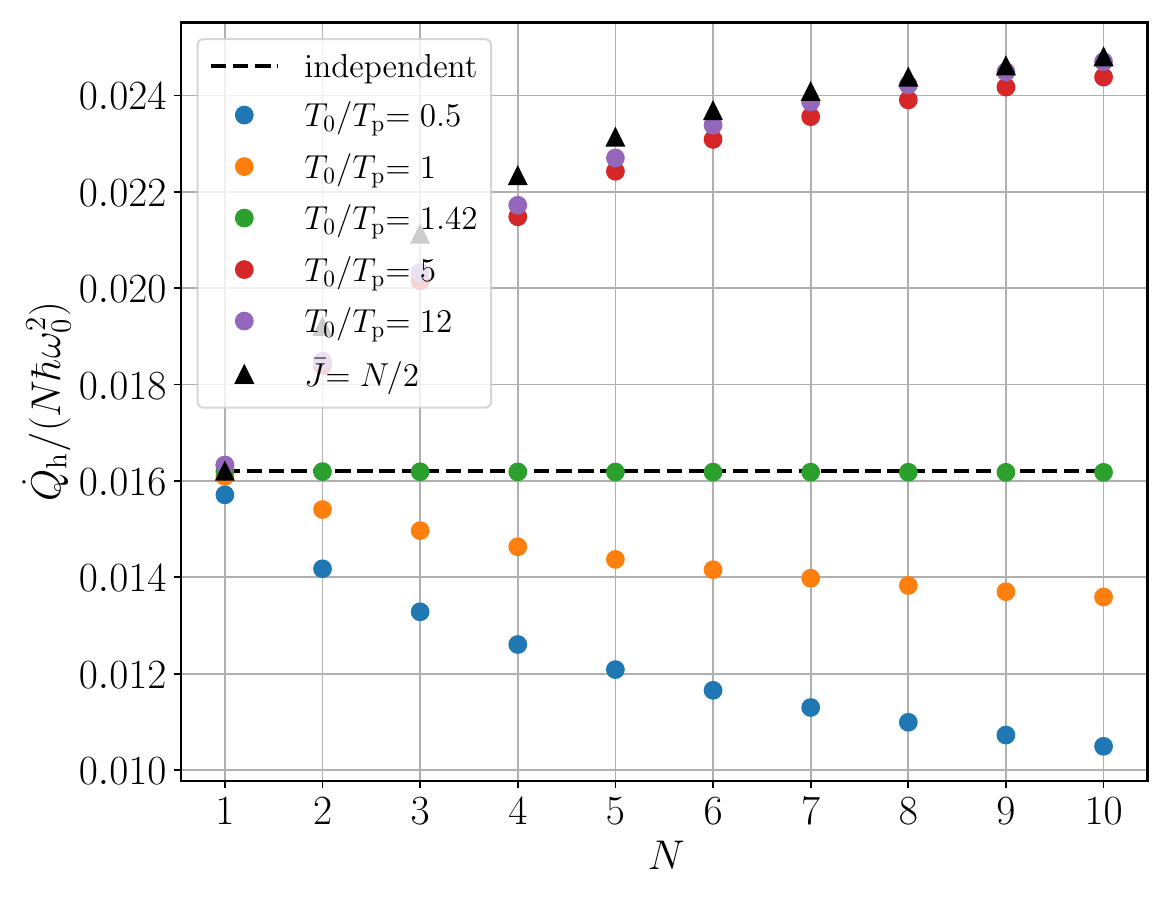}
    \put(0,73){(a)}
  \end{overpic}
\end{minipage}%
\begin{minipage}{.49\textwidth}
  \centering
  \begin{overpic}[width=\linewidth]{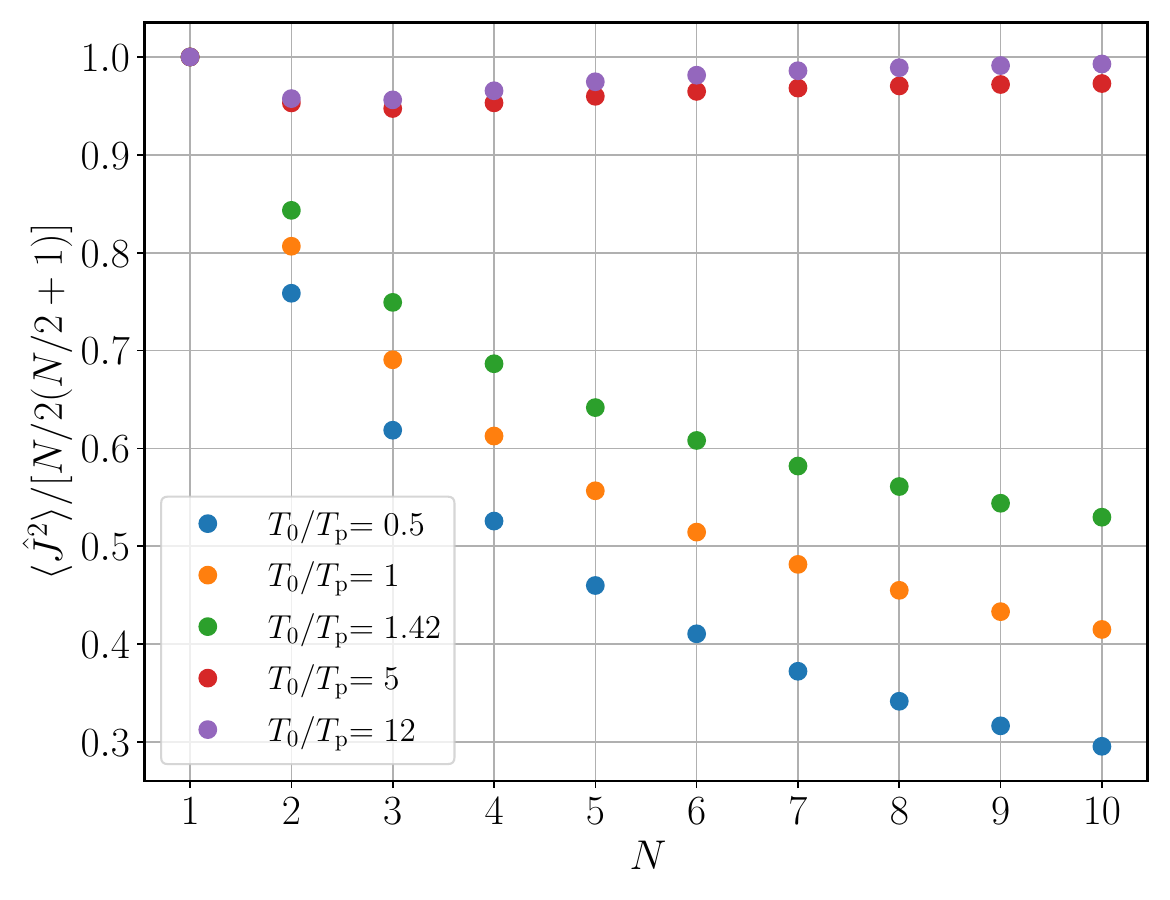}
    \put(0,73){(b)}
  \end{overpic}
\end{minipage}


\begin{minipage}{.49\textwidth}
  \centering
  \begin{overpic}[width=\linewidth]{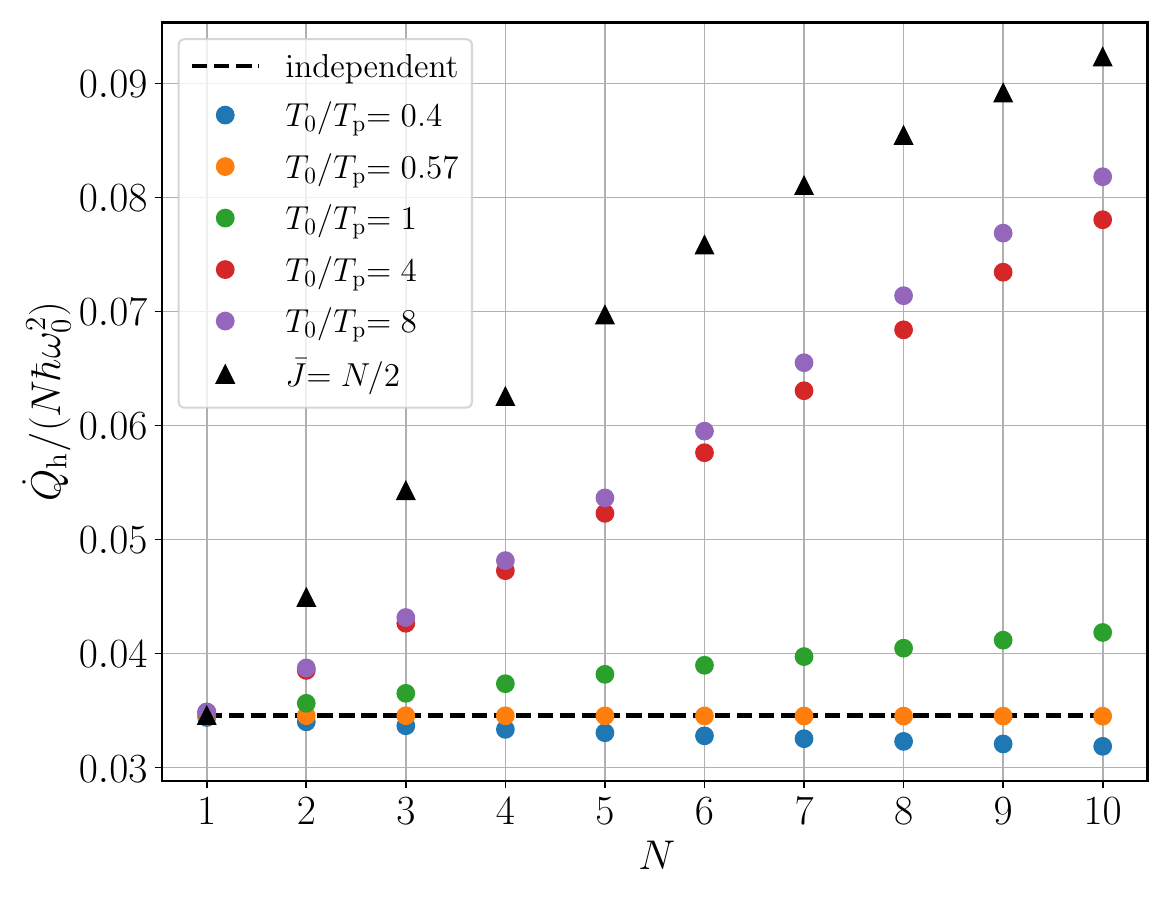}
    \put(0,73){(c)}
  \end{overpic}
\end{minipage}%
\begin{minipage}{.49\textwidth}
  \centering
  \begin{overpic}[width=\linewidth]{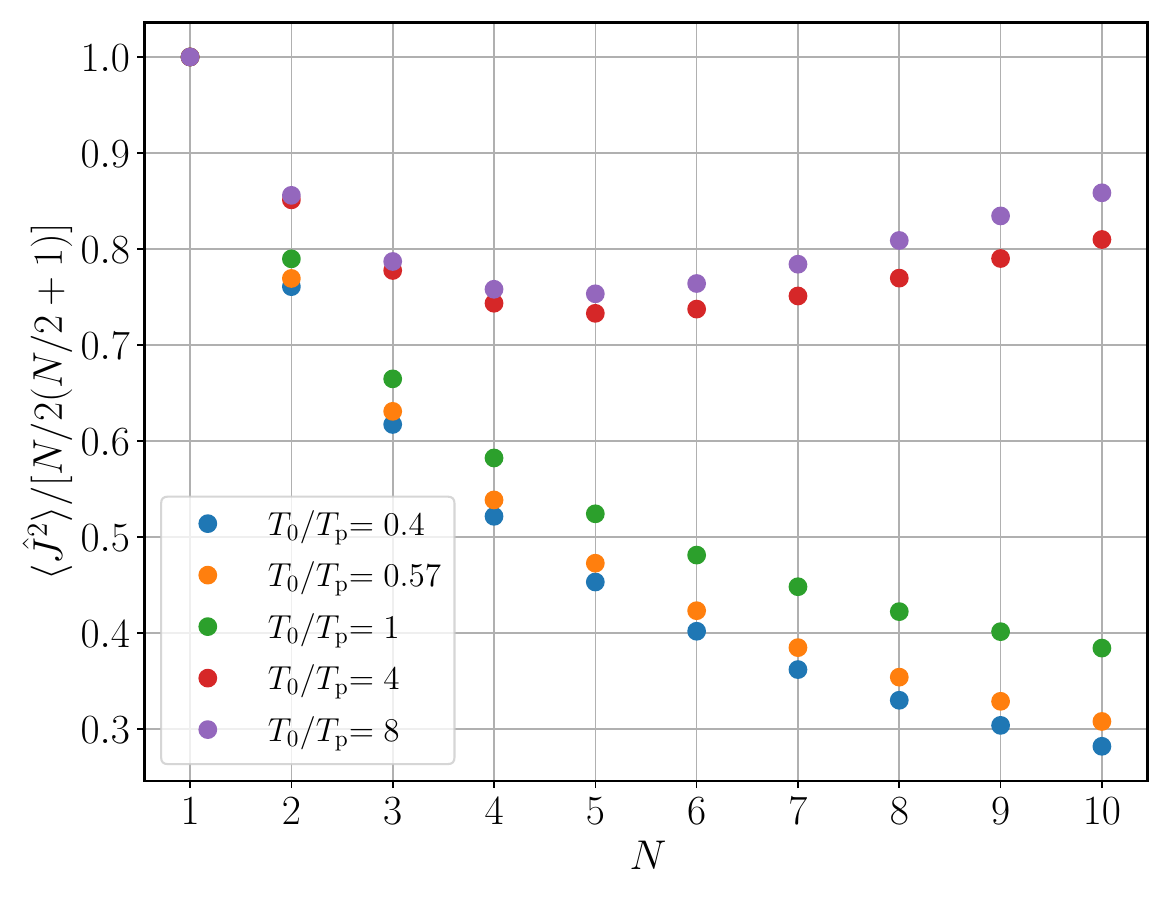}
    \put(0,73){(d)}
  \end{overpic}
\end{minipage}

\caption{(Color online)
Panels (a,c) show the heat current, divided by $N$, as a function of the number of qubits $N$  for various values of the ratio $T_0/T_{\rm p}$. The independent case $\dot{Q}^{\rm ind}_{\rm h}$, given by Eq.~(\ref{eq:current_independent}), is shown as a black dashed line. The ideal case without the parasitic bath, and calculated assuming Eq.~\eqref{eq:Pdelta} with \( \bar{J} = N/2 \), is shown as black triangles.
Panels (b,d) show the total angular momentum $\hat{J}^2$ (normalized by $N/2(N/2+1)$) as a function of  $N$ for various values of the ratio $T_0/T_{\rm p}$.
In panel (a,b) we have set  $T_{\rm h}=T_0$  while $T_{\rm h}=3 T_0$ in panel (c,d). Other parameters are
$T_{\rm c}=T_0/3$, {$\gamma_{\rm h}=\gamma_{\rm c}=\omega_0/10$}, $\gamma_{\rm p}=0.001\omega_0$. }
\label{fig3}
\end{figure*}

In Fig.~\ref{fig3} (a,c) we show how the heat current $\dot{Q}_{\rm h}$, divided by $N$, is influenced by the {\it parasitic} bath.
Panel (a) focuses on the low-temperature regime, i.e. when both primary thermal baths have temperatures lower than \( T_0 \), with \( T_0/T_{\rm h} = 1 \) and \( T_0/T_{\rm c} = 3 \). This corresponds to the system being effectively colder than $T_0$, \( T_0/T^* \approx 1.42\). As we can see, a hot parasitic bath $T_0/T_{\rm p}\leq 1$ (blue and orange dots) suppresses the heat current even below the independent case in Eq.~\eqref{eq:current_independent} (plotted as a black dashed line).
In the particular case where \( T_{\rm p} \approx T^* \approx 1.42 T_0 \) (green dots), the heat current coincides with the value derived for independent qubits, according to Eq.~\eqref{eq:current_independent}. Under this condition, the parasitic bath has the only effect of suppressing the coherence between different qubits, rendering them effectively independent. 
However, when the parasitic bath is sufficiently cold $T_0/T_{\rm p}\geq 10$ (red and purple dots), the heat current $\dot{Q}^{\rm ind}_{\rm h}$ coincides with the one observed when the dynamics are constrained to a single subspace with \( J = N/2 \) ( represented by black triangles). 
Even in this single subspace, a super-extensive scaling in $N$ is absent. This absence is attributed to the low-temperature regime of the baths, which maintains the system close to its ground state, leading to \( \langle -\hat{J}_z \rangle  \approx N/2 \) in Eq.~\eqref{eq:heatcurrent4}, which corresponds to a linear behavior.

Panel (c) illustrates the high-temperature scenario where \( T_0/T_{\rm h} = 1/3 \) and \( T_0/T_{\rm c} = 3 \). In this case, the system has an effective temperature bigger than $T_0$, \( T_0/T^*\approx 0.57 \).  
Similar to the case in panel (a), a hot parasitic bath (blue dots) negatively impacts the heat current.
If \( T_{\rm p} \approx T^* \approx 0.57 T_0 \) (orange dots), the heat current reverts to the value computed for independent atoms using Eq.~\eqref{eq:current_independent}.
However, when the parasitic bath is sufficiently cold (\( T_0/T_{\rm p} \geq 3 \), red and purple dots), the heat current exhibits super-extensive scaling with \( N \). 

Despite this, a discernible gap remains between the heat current \( \dot{Q}_{\rm h} \) calculated in the presence of the parasitic bath, and the ideal one without the parasitic bath calculated assuming Eq.~\eqref{eq:Pdelta} with \( \bar{J} = N/2 \) (represented by black triangles).
This gap exists because the parasitic bath redistributes the population across various subspaces characterized by different values of \( J \), whereas in Eq.~(\ref{eq:Pdelta}) we assume that a single value of $\bar{J}=N/2$ is occupied.

To further interpret our results, in Fig.~\ref{fig3} (b,d) we show the dependence of the total angular momentum $\langle\hat{J}^2\rangle$, divided by $\sim N^2$, on the temperature of the {\it parasitic} bath.
In both panels, we see that a hot parasitic bath (blue, orange, and green dots) corresponds to low values of $\langle\hat{J}^2\rangle$, which is consistent with low values of the heat currents shown in the corresponding panels (a) and (c).
This can be interpreted in the following way. Let us consider the high-temperature regime of the main baths, which is the one that can exhibit the super-linear scaling of the heat current (see Sec.~\ref{ss:fixedsub}). In the limit of infinitely hot parasitic bath, all states would be equally populated. As $N$ increases, the states with $J\sim0$ exponentially outnumber the other states, leading to the observed decrease of $\hat{J}^2$. This, in turn, implies a vanishing $\langle \hat{J}_z\rangle$, which, using Eq.~(\ref{eq:heatcurrent4}), implies a vanishing heat currents even below the independent qubits case - as observed e.g. in the orange and blue dots in panel (a) and (c).

Therefore, to achieve super-extensive scaling, a cold parasitic bath \(T_{\rm p} \ll T_0\) is required to ensure that the subspace with maximum collective spin \(J \approx N/2\) is occupied, corresponding to $\langle\hat{J}^2\rangle\approx N/2(N/2+1)$ in Fig.~\ref{fig3} (b,d). However, this requirement appears to conflict with the need for a high effective temperature $T^*$ to have the super-linear scaling emerging from Eq.~\eqref{eq:Q_steady_limit2}.
Nevertheless, given that the parasitic bath is weakly coupled to the system, it does not significantly influence the system's effective temperature $T^*$, which is dominated by $T_\text{h}$ and $T_\text{c}$. As a result, a cold parasitic bath $T_{\rm p}\ll T_0$ does not necessarily force the system toward its ground state, allowing for a finite window where super-extensive scaling of the heat current can be observed when $T^*\gg T_0$, as shown in Fig.~\ref{fig3} (b).

\section{Experimental proposal}
\label{sect:EP}

\begin{figure}[t]
\centering
 \begin{overpic}[width=0.95\columnwidth]{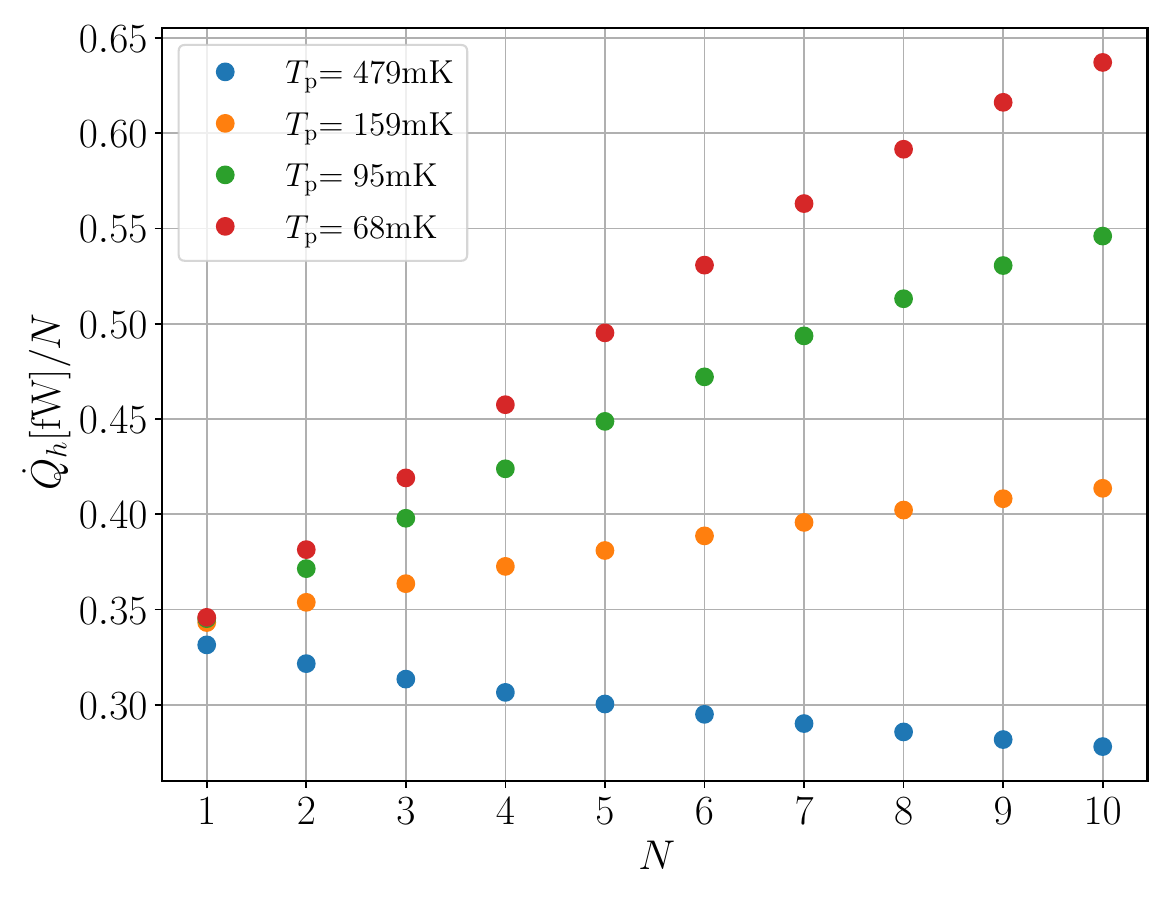}
 \put(0,73){(a)}
 \end{overpic}
 \begin{overpic}[width=0.95\columnwidth]{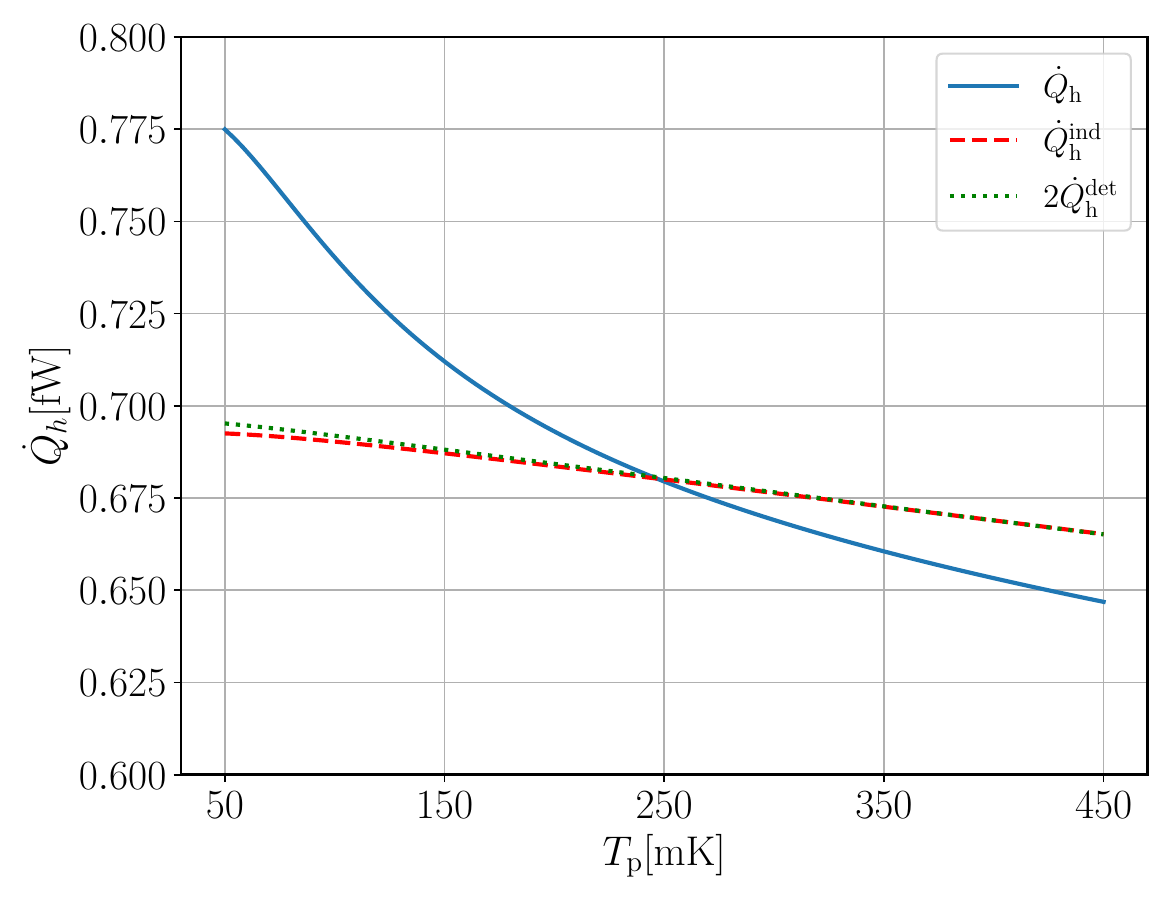}
 \put(0,73){(b)}
 \end{overpic}
\caption{(Color online) Panel (a) shows the heat current $\dot{Q}_{\rm h}$ (divided by $N$) and expressed in femto Watt (fW) as a function of $N$, for different values of $T_{\rm p}$ in milliKelvin (mK).
Here, we choose $\omega_0/(2\pi) = 4.0~{\rm GHz}$, $\gamma_{\rm h} =\gamma_c = 0.5~{\rm GHz}$. Panel (b) shows the heat current $\dot{Q}_{\rm h}$  in femto Watt (fW) as a function of $T_{\rm p}$ in milliKelvin (mK) for $N=2$ qubits.
In this panel, the collective heat current $\dot{Q}_{\rm h}$ is depicted by a blue continuous line, the heat current in the independent scenario $\dot{Q}^{\rm ind}_{\rm h}$ is represented by a red dashed line, and the heat current in the detuned scenario is indicated by a green dotted line. For the detuned scenario, the parameters are set as $\omega^{(1)}_0/(2\pi) = 4.0~{\rm GHz}$ and $\omega^{(2)}_0/(2\pi) = 2.0~{\rm GHz}$, with $\gamma_{1} = 0.5~{\rm GHz}$. The rate $\gamma_2$ is calculated in accordance with Eq.~\eqref{eq:gamma1gamma2}, assuming $\mathcal{Q}_i=20$. In both panels, additional parameters are $\nu_{T_{\rm h}}=8.0~{\rm GHz}$, $\nu_{T_{\rm c}}=1.0~{\rm GHz}$ (corresponding to bath temperatures $T_{\rm h}=384.0~{\rm mK}$ and $T_{\rm c}=96.0~{\rm mK}$), and a parasitic bath coupling rate of $\gamma_{\rm p} = 0.01~{\rm GHz}$.
}
\label{fig2}
\end{figure}

In this Section, we discuss an experimental proposal to observe the enhanced heat current in a circuit QED platform. In Fig.~\ref{fig2} (a) we plot the dependence of the heat current $\dot{Q}_{\rm h}$ on the number of qubits for different temperatures of the parasitic bath, as in Fig.~\ref{fig3}, but using physical parameters - reported in Tab.~\ref{tab:parameters} - that have been used to describe a previous experiment \cite{Ronzani_NatPhys_2018}. This plot demonstrates that the collective enhancement of the heat current is quite pronounced when using parameters that are feasible in experimental settings as long as the temperature of the parasitic bath is roughly below $150\text{mK}$.

\begin{table}[h!]
    \centering
    \begin{tabular}{|c|c|c|c|c|c|c|c|}
   
        \hline
        & \( \omega_0/(2\pi) \) & \( T_{\text{c}} \) & \( T_{\text{h}} \) & \( T_{\text{p}} \) & \( \gamma_{\text{c}} \) & \( \gamma_{\text{h}} \) & \( \gamma_{\text{p}} \) \\
        \hline
        GHz & \( 4.0 \) & \( 2.0 \) & \( 8.0 \) & \( 1.04-10.4 \) & \({ 0.5} \) & \( { 0.5} \) & \( 0.01 \) \\
        \hline
        mK & \( - \) & \( 96.0 \) & \( 384.0 \) & \( 50-500 \) & \( - \) & \( - \) & \( - \) \\
        \hline
    \end{tabular}
    \caption{Physical parameters compatible with the experiment in Ref.~\onlinecite{Ronzani_NatPhys_2018}. In the first line, we reported the frequencies $\nu_T$ associated with the relative temperatures $T$ as $\nu_T\equiv k_{\rm B}T/h$. }
     \label{tab:parameters}
\end{table}

Furthermore, we propose a minimal single-device experiment to demonstrate the collective advantage in a setup with $N=2$ transmon qubits. Ideally, we would like to compare the collective current $\dot{Q}_{\rm h}$ with the independent case $\dot{Q}^{\rm ind}_{\rm h}$ in Eq.~\eqref{eq:current_independent}, where each qubit is independently coupled with its bath. However, working with two different devices introduces variability in Hamiltonian parameters due to fabrication differences which makes a direct comparison difficult. To circumvent this problem, we devise a protocol where we exploit the possibility of controlling the frequency of each transmon by adjusting the external magnetic flux threading through it. By controlling the detuning between the qubits, we can switch between the collective and the independent coupling scheme on the same device. 

Let us assume that the two transmon qubits have tunable frequencies, $\omega_0^{(1)}$ and $\omega_0^{(2)}$. When both of these frequencies are resonant with the $LC$ resonator frequency ($\omega_0^{(k)}=\omega_{ LC}$ for $k=1,2$), the transmons are collectively coupled to the same modes of the baths. Under this condition, the heat current is the collective one $\dot{Q}_{\rm h}$ determined by Eq.~\eqref{eq:heatcurrent3} and discussed in previous sections.
However, if we detune e.g. the second qubit setting $\omega_0^{(2)}=\omega_{\rm LC}/2$, while keeping the first qubit on resonance $\omega_0^{(1)}=\omega_{ LC}$, we can effectively ``disconnect'' the second qubit.
We label the heat current in this configuration as $\dot{Q}^{\rm det}_{\rm h}$. In this detuned scenario, the qubits interact with independent modes of the baths and behave independently. Additionally, due to the significant detuning of the second qubit, it contributes negligibly to the heat current. Therefore, only the first qubit contributes significantly to the heat current, hence we expect that $\dot{Q}^{\rm det}_{\rm h}\approx \dot{Q}^{\rm ind}_{\rm h}/2$.
In Appendix~\ref{appendixHeatDetuned}, we detail the calculation of the heat current $\dot{Q}^{\rm det}_{\rm h}$ in the detuned scenario. This derivation allows us to quantitatively support the previous analysis. In Fig.~\ref{fig2} (b), we show the heat current as a function of the temperature of the parasitic bath for the minimal single-device system comprising $N=2$ transmon qubits. The continuous blue line represents the collective heat current $\dot{Q}_{\rm h}$, while the red dashed line indicates the heat current in the independent case $\dot{Q}^{\rm ind}_{\rm h}$, calculated as per Eq.~\eqref{eq:current_independent}. The green dotted line depicts twice the heat current in the scenario where one qubit is in resonance and the other is detuned, namely $2\dot{Q}^{\rm det}_{\rm h}$ and determined by Eq.~\eqref{eq:heatcurrent_detuned}. As we can see in the figure, the detuned case can be used to accurately estimate the independent scenario as  $2\dot{Q}^{\rm det}_{\rm h}\approx \dot{Q}^{\rm ind}_{\rm h}$, and thus to experimentally validate and quantify the heat current enhancement in the collective case.

Significantly, Fig.~\ref{fig2} (b) reveals that at a parasitic bath temperature of $T_{\rm p}=50$ mK, the collective $\dot{Q}_{\rm h}$ exhibits an enhancement of approximately $13\%$ compared to the independent scenario. In contrast, at $T_{\rm p}=450$ mK, the collective heat current shows a reduction of about $4\%$ relative to the independent case.
Concerning the experimental implementation, fabricating a multi-qubit circuit is not significantly more demanding than making a single qubit. We believe that the main experimental challenge in realizing such a minimal single-device experiment is ensuring that the two transmons have the same frequency $\omega_0$ and are coupled to the same bath with the same intensity to guarantee the conservation of $\hat{J}^2$. Small imperfections in the physical implementation would inevitably result in breaking the conservation of $\hat{J}^2$. However, heuristically, these effects would be similar to the presence of the parasitic baths that we studied earlier.

\section{Discussion and conclusions}
\label{sec:Conclusions}

In this work, we have illustrated how the collective interactions between $N$ qubits and two thermal baths can enhance the heat transport across the device, 
compared to a scenario where each qubit is coupled independently with the baths.

The device can be implemented within current quantum technologies, and we focus on its circuit-QED implementation.
After revealing the physical origin of the collective enhancement of the heat current, we analyze its resilience to a third, parasitic thermal bath that acts locally on each qubit, modeling unavoidable single-qubit noise.
We then assess the experimental feasibility of our proposal by choosing system parameters that have been measured in a previous experiment \cite{Ronzani_NatPhys_2018}, and we propose a minimal experimental device based on two superconducting qubits.
 Our findings indicate that this collective advantage is not only robust to local noise but also observable under experimental conditions.

It is important to note that the circuit depicted in Fig.~\ref{fig:sketch} is a simplified model. In real-world scenarios, additional factors may come into play. For example, unintended capacitive coupling between qubits is likely to occur in practical implementations, leading to effective dipole-dipole interactions between qubits. However, this coupling would primarily affect the conservation of $\hat{J}^2$, leading to effects qualitatively analogous to those introduced by the parasitic bath.

Another important limitation of our predictions stems from the use of a Markovian Lindblad master equation, which is well justified only for small transition rates compared to the qubit frequency $\omega_0$. This is primarily required by the secular approximation and the weak coupling approximation (commonly referred to as the Born-Markov approximation), as detailed in the derivation of our master equation in Appendix~\ref{App:derivation_Lindblad}. Given that we chose relatively small rates, approximately $10\%$ of the bare frequency $\omega_0$, the use of the Lindblad master equation is reasonably justified. Additionally, it has been shown that non-secular terms tend to be more relevant during transient dynamics, while their influence is significantly reduced at the steady state \cite{Farina22}. Nonetheless, it would be interesting to explore the influence of non-secular terms and non-Markovian effects that emerge at strong coupling and can play a significant role in an experiment \cite{Meng21}.

In the future, it will be interesting to investigate similar collective effects when time-dependent driving is present allowing the device to function also as a heat engine \cite{Quan_PhysRevE_2007,Pietzonka_PRL_2018} or as a refrigerator \cite{Linden_PRL_2010,Mari_PRL_2012}, depending on the specific physical parameters.

\section{Acknowledgments}

Numerical work has been performed by using the Python toolbox
QuTiP2~\cite{QuTip} and the PIQS library~\cite{Shammah_PhysRevA_2018}. We wish to thank R. Fazio, F. Campaioli, D. De Bernardis, L. Giacomelli, M. Brunelli, D. Ferraro, and F. Minganti for useful discussions. G.M.A and M.S. acknowledge funding from the European Research Council (ERC) under the European Union's Horizon 2020 research and innovation program (Grant agreement No. 101002955 -- CONQUER).
P. A. E. gratefully acknowledges
funding by the Berlin Mathematics Center MATH+
(AA2-18).  F.N gratefully acknowledges funding by the BMBF (Berlin Institute for the Foundations of Learning and Data -- BIFOLD), the European Research Commission (ERC CoG 772230) and the Berlin Mathematics Center MATH+ (AA1-6, AA2-8). This work was supported by Research Council of Finland
Grant No. 312057 (Centre of Excellence program).

\appendix 
\section{Derivation of the Hamiltonian \eqref{eq:H_tot}}
\label{App:derivation_of_TC}

In this Section, we provided a detailed derivation of the Hamiltonian in Eq.~\eqref{eq:H_tot}.
We begin by revisiting the quantization processes for two key components of the circuits drawn Fig.~\ref{fig:sketch}(b): the $LC$ circuit and the transmon qubit. The derivation of these Hamiltonians is detailed in Subsections \ref{ss:LC} and \ref{ss:transmon}, respectively.
Building upon these concepts, we then proceed to derive the Hamiltonian in Eq.~\eqref{eq:H_tot} from the circuit drawn Fig.~\ref{fig:sketch}.

\subsection{Quantization of the $LC$ superconducting circuit}
\label{ss:LC}

The Hamiltonian for an $LC$ circuit, where \( L \) is the inductance and \( C \) is the capacitance, is given by:

\begin{equation} \label{eq:hamiltonian_classical}
\hat{H}_{LC} = \frac{\hat{Q}^2}{2C} + \frac{\hat{\phi}^2}{2L}~,
\end{equation}
where \( \hat{Q} \) denotes the charge on the capacitor and \( \hat{\phi} \) represents the magnetic flux through the inductor. These two variables fulfill the canonical commutation relation $[{\hat{Q}}, {\hat{\phi}}] = i\hbar$~.
Being Eq.~\eqref{eq:hamiltonian_classical} a quadratic Hamiltonian, it is useful to express \( \hat{Q} \) and \( \hat{\phi} \) in terms of ladder operators
$\hat{a}, \hat{a}^\dagger$ as

\begin{equation} \label{eq:Q_expression}
\hat{Q} = \sqrt{\frac{\hbar}{2Z_{ LC}}} (\hat{a} + \hat{a}^\dagger)~,
\end{equation}
\begin{equation} \label{eq:Phi_expression}
\hat{\phi} = i\sqrt{\frac{\hbar Z_{ LC}}{2}} (\hat{a}^\dagger - \hat{a})~,
\end{equation}
where ladder operator statisfy $[{\hat{a}}, {\hat{a}}^\dagger] = 1$ and $Z_{\rm LC}=\sqrt{L/C}$ is the impedance associated with the $LC$ circuit.
The Hamiltonian can be thus diagonalized in the form 
\begin{equation} \label{eq:hamiltonian_LC}
\hat{H}_\text{LC} = \hbar \omega_{ LC} \left( {\hat{a}}^\dagger {\hat{a}} + \frac{1}{2} \right)~,
\end{equation}
where the resonant frequency \( \omega_{LC} \) of the $LC$ circuit reads $\omega_{ LC} = {1}/{\sqrt{LC}}$.

\subsection{Quantization for a transmon qubit}
\label{ss:transmon}

A transmon qubit can modeled as a nonlinear circuit composed of a shunt capacitance and a Josephson junction in parallel. The Hamiltonian for the transmon qubit can be written as:

\begin{equation} \label{eq:transmon_full_hamiltonian}
\hat{H}_{\text{T}} = \frac{1}{2C_{\rm T}} \hat{Q}^2_{\rm T} - E_\text{J} \cos\left(\frac{2\pi\hat{\phi}_{\rm T}}{\phi_0}\right)~,
\end{equation}
where $C_{\rm T}$ is the transmon capacitor, $\phi_0=h/(2e)$ and \( E_\text{J} \) represents the Josephson energy.
The variables \( \hat{Q}_{\rm T} \) and \(  \hat{\phi}_{\rm T} \) are quantum operators representing the charge of excess Cooper pairs and the phase difference across the Josephson junction, respectively. Again, they satisfy the canonical commutation relation:

\begin{equation} \label{eq:commutation_n_phi}
[\hat{Q}_{\rm T}, \hat{\phi}_{\rm T}] = i{\hbar}~.
\end{equation}
Trasmon qubits operate in the regime \( E_\text{J} \gg E_C \) - where \( E_C={e^2}/({2 C})\) denotes the charging energy - to minimize sensitivity to charge noise.  When the potential of the transmon is examined in this regime, it can be approximated as a quasi-harmonic potential. 
Linearizing around the minimum of this potential, the cosine term in the Hamiltonian can be expanded, leading to an almost harmonic behavior with small anharmonic corrections. 

Neglecting the non-linearity stemming from the Josephson energy, Eq.~\eqref{eq:transmon_full_hamiltonian} can be approximated as

\begin{equation} \label{eq:simplified_harmonic}
\hat{H}^\prime_{\text{T}} = \frac{1}{2C_{\rm T}} \hat{Q}^2_{\rm T}  + \frac{1}{2L_\text{J}} \hat{\phi}^2_{\rm T}~,
\end{equation}
where $L_\text{J}=(\hbar/2e)^2(1/E_\text{J})$.

This is analogous to the Hamiltonian of a quantum harmonic oscillator in Eq.~\eqref{eq:hamiltonian_classical}. Hence,  \( \hat{Q}_{\rm T} \) and \( \hat{\phi}_{\rm T} \) can be expressed in terms of ladder operators $\hat{b}, \hat{b}^\dagger$ (fulfilling the commutation relation $[\hat{b}, \hat{b}^\dagger] = 1$ ) as

\begin{equation} \label{eq:Q_expressionT}
\hat{Q}_{\rm T} = \sqrt{\frac{\hbar}{2Z_{\rm T}}} (\hat{b} + \hat{b}^\dagger)~,
\end{equation}
\begin{equation} \label{eq:Phi_expressionT}
\hat{\phi}_{\rm T} = i\sqrt{\frac{\hbar Z_{\rm T}}{2}} (\hat{b}^\dagger - \hat{b})~,
\end{equation}
where $Z_{\rm T}=\sqrt{L_{\rm J}/C_{\rm T}}$ is the impedance associated with the equivalent $LC$ circuit.
Substituting Eqs.~\eqref{eq:Q_expressionT} and \eqref{eq:Phi_expressionT} into the linearized Hamiltonian $\hat{H}^\prime_{\text{T}}$ we have
\begin{equation} \label{eq:harmonic_ladder}
\hat{H}^\prime_{\text{T}} = E^\prime_\text{T} \hat{b}^\dagger \hat{b}~.
\end{equation}
where \( E^\prime_\text{T}=\sqrt{8E_\text{J}E_{C}} \) is the effective frequency of the transmon.
This is the harmonic spectrum of the transmon, at the lowest order in $E_C/{E_\text{J}}$.
However, the presence of anharmonic terms is crucial for the transmon to function as a qubit. Specifically, the anharmonicity ensures that the energy levels of the transmon are not evenly spaced, allowing for selective addressing of specific energy transitions.

The quartic anharmonic term in the potential, which arises from the Taylor expansion of the cosine function in Eq.~\eqref{eq:transmon_full_hamiltonian}, can be expressed as

\begin{equation} \label{eq:quartic_potential}
\delta \hat{H}_\text{T}= -\frac{E_J}{4!} \left(\frac{2\pi\hat{\phi}_{\rm T}}{\phi_0}\right)^4~,
\end{equation}
where the transmon Hamiltonian is approximated by  $\hat{H}_\text{T}\approx \hat{H}^\prime_\text{T}+\delta \hat{H}_\text{T}$.
Given the relationship between \( \hat{\phi} \) and \( \hat{b} \) and \( \hat{b}^\dagger \) in Eq.~\eqref{eq:Phi_expressionT} the quartic term can be re-expressed in terms of these ladder operators.
Expanding \( \hat{\phi}^4 \) and neglecting off-diagonal terms that do not conserve the number of excitations, the perturbation $\delta \hat{H}_\text{T}$ becomes:

\begin{equation} \label{eq:perturbation_ladder}
\delta \hat{H}_\text{T} = -\frac{E_C}{2} (\hat{b}^\dagger \hat{b}^\dagger \hat{b} \hat{b} + 2\hat{b}^\dagger \hat{b})~.
\end{equation}

The term \( \hat{b}^\dagger \hat{b}^\dagger \hat{b} \hat{b} \) in the potential introduces an anharmonicity, leading to unevenly spaced energy levels. This anharmonicity -proportional to $E_C$- is crucial for the operation of the transmon as a qubit, allowing for selective addressing of its states. The term \( \hat{b}^\dagger \hat{b} \) results in a small shift in the effective energy \( E_\text{T} \) of the transmon

\begin{equation} \label{eq:effective_frequency}
E_\text{T} \approx \sqrt{8E_\text{J} E_C} - {E_C}~.
\end{equation}

This equation incorporates the anharmonic corrections from the Josephson potential.

Due to the introduced anharmonicity, our focus can be limited to the subspace spanned by the unexcited state $\ket{0}_{\rm T}$ (where $\hat{b}\ket{0}_{\rm T}=0$) and the excited state,  $\ket{1}_{\rm T}\equiv\hat{b}^\dagger\ket{0}_{\rm T}$.
 In this subspace, we can truncate ladder operator to $\hat{b}\approx\hat{\sigma}_{-}$, $\hat{b}^\dagger\approx\hat{\sigma}_{+}$ and subsequently $\hat{b}^\dagger\hat{b}\approx (\hat{\sigma}_z+1)/2$.
Hence, transmon's observables can be  expressed in the truncated basis as:

\begin{align}
\label{eq:truncated}
\hat{H}_\text{T}=\frac{E_\text{T}}{2}(\hat{\sigma}_z+1)~,\\
\hat{Q}_{\rm T} = \sqrt{\frac{\hbar}{2Z_{\rm T}}} \hat{\sigma}_x~,\\
\hat{\phi}_{\rm T} = - \sqrt{\frac{\hbar Z_{\rm T}}{2}}\hat{\sigma}_y~.
\end{align}

\subsection{Derivation of the Hamiltonian of the system}
\label{ss:derivation}

Here we derive the Hamiltonian of the system introduced in the main text. The lumped-element circuit diagram, including capacitances, inductances, resistances, and the various variables is depicted in Fig.~\ref{fig:sketch}.  This circuit corresponds to the following Lagrangian for the transmons
\begin{equation}
\begin{split}
\label{eq:Ltot}
  \mathcal{L} &= \frac{1}{2}\sum_{j=1}^N\left[C_{\rm T}\dot{\phi}_j^2- \frac{1}{L_{\rm J}}{\phi}_j^2+\sum_{i={\rm h,c}}C^{(i)}_{\rm c}(\dot{\phi}_j-V_i)^2\right] ~, 
\end{split}
\end{equation}

where  $V_i$ is the voltage drop occurring in the resistance $R_i$ and $C^{(i)}_{\rm c}$ corresponds to the capacitive coupling between the transmons and the $RLC$ circuits. In the following, the capacitive couplings are assumed to be small with respect to the transmon capacities, $C^{(i)}_{\rm c}\ll C_{\rm T}$.  Under this approximation, in the following, we will trace out the variables $V_i$ and obtain a Markovian master equation for the transmons.
The Lagrangian gives rise to canonical momenta \( Q_j\) (the charges), which are the derivatives of the Lagrangian with respect to \( \dot{\phi}_j \), $Q_j\equiv \partial\mathcal{L} /\partial\dot{\phi_j}$. In terms of $\dot{\phi}_j$ the charges \( Q_j\) can be expressed as: 
\begin{equation}
\begin{split}
Q_j=(C_{\rm T}+\sum_{i={\rm h,c}}C^{(i)}_{\rm c})\dot{\phi}_j-\sum_{i={\rm h,c}}C^{(i)}_{\rm c}V_i~, 
\end{split}
\end{equation}
The previous equation can be inverted to obtain the voltages $\dot{\phi}_j$. This relation, at the first order in $C^{(i)}_{\rm c}/C_{\rm T}$, reads

\begin{equation}
\begin{split}
\dot{\phi}_j=Q_j\left(\frac{1}{C_{\rm T}}-\frac{\sum_{i={\rm h,c}}C^{(i)}_{\rm c}}{C^2_{\rm T}}\right)+\sum_{i={\rm h,c}}C^{(i)}_{\rm c}V_i~, 
\end{split}
\end{equation}

The Hamiltonian of the system can be thus obtained as the Legendre transform of the total Lagrangian $\mathcal{L}$ in Eq.~\eqref{eq:Ltot} as $H \equiv \sum_{j=1}^N {Q}_j\dot{{\phi}}_j-\mathcal{L}$. At this stage is now possible to quantize the system Hamiltonian by promoting classical variables to quantum ones as follows,

\begin{equation}
\begin{split}
\phi_j&\to \hat{\phi}_j~,\\
Q_j&\to \hat{Q}_j~,\\
V_i&\to \hat{V}_i~,\\
\end{split}
\end{equation}

The total quantized Hamiltonian can be expressed at the first order $C^{(i)}_{\rm c}/C_{\rm T}$ as:

\begin{equation}
\begin{split}
\label{eq:H_circuit}
\hat{H} &=\sum_{j=1}^N\frac{1}{2}\left(\frac{1}{C_{\rm T}}\left[1-\frac{\sum_{i={\rm h,c}}C^{(i)}_{\rm c}}{C_{\rm T}}\right] \hat{Q}_j^2  + \frac{1}{L_\text{J}} \hat{\phi}^2_j\right)+\\&-\sum_{j=1}^N\sum_{i={\rm h,c}}\frac{C^{(i)}_{\rm c}}{C_{\rm T}}\hat{Q}_j\hat{V}_i~, 
\end{split}
\end{equation} 
where we neglected terms $\hat{V}^2_i$ that do not act on the transmons.
By truncating this Hamiltonian on the transmon qubit Hilbert space, as in Eq.~\eqref{eq:truncated}, we have:
\begin{equation}
\begin{split}
\hat{H} &\approx\sum_{j=1}^N \frac{E_\text{T}}{2}(\hat{\sigma}^{(j)}_z+1)
-\sum_{j=1}^N\sum_{i={\rm h,c}}{\hbar G_i}\frac{ \hat{\sigma}^{(j)}_x}{2}\hat{V}_i~, 
\end{split}
\end{equation} 

where coupling between the qubits and the voltage $\hat{V}_i$ is denoted by $G_i$, which is defined as $G_i=2({C^{(i)}_{\rm c}}/{C_{\rm T}})(2/\hbar Z_{\rm T})^{1/2}$. In the main text, we have used the notation $\hbar\omega_0=E_{\rm T}$ to represent the qubit energy. With these definitions and notations in place, we have completed the derivation of the Hamiltonian of the system, given in Eq.~\eqref{eq:H_tot_J}.

\section{Derivation of Lindblad master equation}
\label{App:derivation_Lindblad}

In this Section, we derive the Lindblad master equation for the collection of transmons by tracing out the $RLC$ degrees of freedom. The complete system is described by the total Hamiltonian \(\hat{H}_{\text{tot}}\) :

\begin{equation}
\label{eq:Htot}
\hat{H}_{\text{tot}} = \hat{H} + \hat{H}_{{RLC}}~,
\end{equation}

 Here, \(\hat{H}_{{RLC}}\) represents the Hamiltonian for $RLC$ circuits while $\hat{H}$ is the transmons Hamiltonian defined in Eq.~\eqref{eq:H_tot_J}. As \(\hat{H}_{{RLC}}\) contains a resistive part, writing an explicit expression involves the coupling with a collection of infinite $LC$ elements, as discussed for example in Ref.~\onlinecite{Cattaneo_AdvQuantumTech_2021}. As we will show later, here, we do not need to write down the full $RLC$ Hamiltonian, \(\hat{H}_{{RLC}}\). Instead, thanks to the fluctuation-dissipation theorem, it will be sufficient to know the classical equation of motion of the $RLC$ circuit.   We further define the interaction Hamiltonian \(\hat{H}_{\text{int}}\) as:

\begin{align} 
\label{eq:H_int_app}
\hat{H}_{\text{int}}&=\hat{H}_{\text{tot}}-\hat{H}-\hat{H}_{{RLC}}~,\\&=
-\sum_{i={\rm h,c}}{\hbar G_i}\hat{J}_x \hat{V}_i~,
\end{align}

We employ the Born-Markov approximation, which assumes that the qubits are weakly coupled to the modes of the baths, to trace out the field and obtain a master equation governing the evolution of the qubits' density matrix \(\rho(t)\). 
The calculations are facilitated by working in the interaction picture. For any generic operator \(\hat{O}\), its corresponding form in the interaction picture is \(\hat{O}^{\text{I}}(t) = U_0^\dagger(t) \hat{O} U_0(t)\), where \(U_0(t) = \exp[-i \hat{H}_{\rm bare} t]\) and \(\hat{H}_{\rm bare} = \hat{H}_0 + \hat{H}_{\text{r}}\). Meanwhile, the density matrix in the interaction picture evolves according to \(\hat{\rho}^{\text{I}}(t) = U_0(t) \hat{\rho}(t) U_0^\dagger(t)\).

The evolution of the total density matrix in the interaction picture, \( \hat{\rho}^{\text{I}}_{\text{tot}}(t) \), is dictated by the following master equation~\cite{Petruccione}:

\begin{equation}
\partial_t \hat{\rho}^{\text{I}}_{\text{tot}}(t) = -\frac{1}{\hbar^2} \int_0^\infty d\tau \left[ \hat{H}^{\text{I}}_{\text{int}}(t), \left[ \hat{H}^{\text{I}}_{\text{int}}(t-\tau), \hat{\rho}^{\text{I}}_{\text{tot}}(t) \right] \right].
\end{equation}

 In reaching this equation, we apply the {\it Markov approximation}, assuming that the environment’s correlation time $\tau_{\rm RLC}$  is much shorter than the timescale of the qubit's dynamics, $\omega^{-1}_0$, namely $\omega_0\tau_{\rm RLC}\ll1$. The  

This allows us to approximate \( \hat{\rho}^{\text{I}}_{\text{tot}}(t-\tau) \approx \hat{\rho}^{\text{I}}_{\text{tot}}(t) \) under the time integral and to extend the integration limit to \( \tau \to \infty \).

Upon tracing out the field subsystem, the master equation for the matter density matrix \( \hat{\rho}^{\text{I}}(t) \) is given by:

\begin{align}
\label{eq:master_equation}
\partial_t \hat{\rho}^{\text{I}}(t) &= -\frac{1}{\hbar^2} \int_0^\infty d\tau \Bigg\{ \text{tr}_{\text{r}} \left[ \hat{H}^{\text{I}}_{\text{int}}(t) \hat{H}^{\text{I}}_{\text{int}}(t-\tau) \hat{\rho}^{\text{I}}_{\text{r}}(t) \right] \hat{\rho}^{\text{I}}(t) \nonumber \\
&- \text{tr}_{\text{r}} \left[ \hat{H}^{\text{I}}_{\text{int}}(t) \hat{\rho}^{\text{I}}_{\text{r}}(t) \hat{\rho}^{\text{I}}(t) \hat{H}^{\text{I}}_{\text{int}}(t-\tau) \right] + \text{h.c.} \Bigg\}~,
\end{align}

where $\text{tr}_{\text{r}} [\dots]$ denotes the trace on the $RLC$ degrees of freedom.
Here, we employed the {\it Born approximation}, which assumes that the $RLC$ circuits serve as large memory-less reservoirs. This lets us approximate \( \hat{\rho}^{\text{I}}_{\text{tot}}(t) \approx \hat{\rho}^{\text{I}}(t) \hat{\rho}^{\text{I}}_{\text{r}}(t) \).

 In the interaction picture, the interaction Hamiltonian defined in Eq.~\eqref{eq:H_int_app} reads:
\begin{equation}
 \hat{H}^{\rm I}_{{\rm int}}(t) =-\sum_{i={\rm h,c}} \hbar G_i\left[\hat{J}_-^{\rm I}(t)+ \hat{J}_+^{\rm I}(t)\right]\hat{V}^{\rm I}_i(t)~,
\end{equation}

Here, \( \hat{V}^{\rm I}_i(t) \) are the voltage operators associated with the $RLC$ circuits, and \( \hat{J}_+^{\rm I}(t) \) and \( \hat{J}_-^{\rm I}(t) \) are the qubits collective raising and lowering operators.
The qubit raising and lowering operators in the interaction picture are given by:

\begin{eqnarray}
 \hat{J}_+^{\rm I}(t) &=& e^{i \omega_0 t} \hat{J}_+~,\\
 \hat{J}_-^{\rm I}(t) &=&  e^{-i \omega_0 t}  \hat{J}_-~.   
\end{eqnarray}

The phase factors arise due to the transformation to the interaction picture, where $\omega_0$  is the bare frequency of the qubits.
By substituting ladder operators in the interaction picture into the master equation (Eq.~\eqref{eq:master_equation}) and performing the secular approximation, we select operators that induce transitions between eigenstates. The secular approximation involves neglecting rapidly oscillating terms in the interaction picture, which is valid when \(\omega_0\) is large compared to the system-bath couplings.
In this regime, we obtain:

\begin{widetext}
 \begin{eqnarray}
\label{eq:master_equation1}
\partial_t\hat{\rho}(t) 
&=&-\frac{i}{\hbar}[\hat{H}_0, \hat{\rho}(t)]- \sum_{i,j} {G_iG_j} \Big\{ S_{\hat{V}_i,\hat{V}_j}(-\omega_0)
\big[ \hat{J}_{+} \hat{J}_{-}\hat{\rho}(t) -\hat{J}_{+} \hat{\rho}(t)\hat{J}_{-}\big]+S_{\hat{V}_i,\hat{V}_j}(\omega_0) \big[ \hat{J}_{-} \hat{J}_{+}\hat{\rho}(t) - \hat{J}_{-} \hat{\rho}(t)\hat{J}_{+}\big] +{\rm h.c.}\Big\}~,
\end{eqnarray}   
\end{widetext}
where $S_{\hat{V}_i,\hat{V}_j}(\omega)$  is the voltage dynamical structure factor of the $RLC$ circuits, 
\begin{equation}
\label{eq: voltage dynamical structure factor}
S_{\hat{V}_i,\hat{V}_j}(\omega) \equiv \int_0^\infty d\tau  e^{-i \omega  \tau} \langle \hat{V}_i^{{\rm I}}(\tau)\hat{V}_j\rangle_{\rm r} ~,
\end{equation}

The voltage dynamical structure factor can be calculated employing the fluctuation-dissipation theorem \cite{Cattaneo_AdvQuantumTech_2021}:  

\begin{equation}
\label{eq:FDT}
S_{\hat{V}_i,\hat{V}_j}(\omega)= 2\hbar\omega  {\rm Re} [ Z_{{\rm tot},i}(\omega) ]\mathcal{N}_i(\omega)
\delta_{i,j}~,
\end{equation}
where $ Z_{{\rm tot},i}(\omega)$ is the total impedance of the $RLC$ circuit and 
$\mathcal{N}_i(\omega)=[n_i(\omega)+1]\Theta(\omega)-n_i(-\omega)\Theta(-\omega)$ is a thermal occupation factor.
The total impedance of the $RLC$ circuits in Fig.~\ref{fig:sketch} can be obtained by considering that the $LC$ branch is in parallel with the resistor branch, hence:
\begin{equation}
\label{eq:sumofZ}
    Z_{{\rm tot},i}(\omega)(\omega)=\left(\frac{1}{ Z_{LC}(\omega)}+\frac{1}{Z_{R_i}}\right)^{-1}~,
\end{equation}
where $Z_{R_i}=R_i$ is the impedance of the $i$-th resistor and $Z_{LC}(\omega)$ is the impedance of the $LC$ circuit,

\begin{equation}
    Z_{LC}(\omega)=\frac{i}{C}\frac{\omega}{\omega^2-\omega_{\rm LC}^2}~.
\end{equation}
By placing the explicit form of $Z_{LC}(\omega)$ and $Z_{R_i}$ in Eq.~\eqref{eq:sumofZ} we obtain:

\begin{equation}
  {\rm Re} [  Z_{{\rm tot},i}(\omega)]=\frac{\omega^2R_i}{\omega^2+R_i^2C^2(\omega^2-\omega_{ LC}^2)^2}~,
\end{equation}

This expression can be further simplified by introducing the quality factor $\mathcal{Q}_i$ as the ratio of the frequency of the $RLC$ circuit divided by its line width $\kappa_i$, $\mathcal{Q}_i=\omega_{ LC}/\kappa_i$. The linewidth of the $RLC$ circuit resonance is given by $\kappa_i=R_i$. Hence, we have:

\begin{equation}
  {\rm Re} [  Z_{{\rm tot},i}(\omega)]=\frac{R_i}{1+\mathcal{Q}_i^2(\frac{\omega}{\omega_{ LC}}-\frac{\omega_{ LC}}{\omega})^2}~,
\end{equation}
By plugging this equation in Eq.~\eqref{eq:FDT} we obtain an explicit expression for the dynamical structure factor: 

\begin{equation}
S_{\hat{V}_i,\hat{V}_j}(\omega)=\left[\frac{\ 2\hbar\omega  R_i \mathcal{N}_i(\omega)}{1+\mathcal{Q}_i^2(\frac{\omega}{\omega_{ LC}}-\frac{\omega_{ LC}}{\omega})^2}\right]\delta_{ij} ~.
\end{equation}


By plugging the explicit form of the dynamical structure factor in Eq.~\eqref{eq:master_equation1}, we can finally arrive at a master equation that can be written in the form of Eq.~\eqref{eq:Lindblad}

 \begin{equation}   
\label{eq:master_equation2}
\partial_t\hat{\rho}(t)=-\frac{i}{\hbar}[\hat{H}_0, \hat{\rho}(t)]+\mathcal{D}[\hat{\rho}(t)]~.
 \end{equation}

Here, the dissipator \( \mathcal{D}[\rho(t)] \) is given by:

 \begin{eqnarray}
\label{eq:Dissipator_App}
\mathcal{D}[\rho(t)]&=&  \sum_{i={\rm h,c}}\gamma_i\Bigg[(1 + n_{i}) \left( \hat{J}_- \hat{\rho} \hat{J}_+ - \frac{1}{2} \left\{ \hat{J}_+ \hat{J}_-, \hat{\rho} \right\} \right)+\nonumber\\&+&{n_{i}}\left( \hat{J}_+ \hat{\rho} \hat{J}_- - \frac{1}{2} \left\{ \hat{J}_- \hat{J}_+, \hat{\rho} \right\} \Bigg]\right)~,
\end{eqnarray}
where the rates read

\begin{equation}
\label{eq:rates}
 \gamma_i=  \frac{ 4\hbar\omega_0  R_i G_i^2}{1+\mathcal{Q}_i^2(\frac{\omega_0}{\omega_{ LC}}-\frac{\omega_{ LC}}{\omega_0})^2}~.
\end{equation}

\section{Details on the heat current }
\label{appendixA}

In this Appendix, we provide detailed mathematical calculations related to the heat current. Subsection \ref{ss:derivation20} is dedicated to deriving an explicit expression for the heat current, specifically Eq.~\eqref{eq:heatcurrent3}. Following this, in Subsection \ref{ss:analitycalexpress} we discuss various limits within which a simpler analytical expression for the heat current can be obtained.

\subsection{Derivation of Eq.~\eqref{eq:heatcurrent3}}
\label{ss:derivation20}
Starting from Eq.~\eqref{eq:heatcurrent2}, we can express the heat current contribution from a specific bath.
 We will focus on deriving the heat current from the cold bath, \(\dot{Q}_{\rm h}\).

The heat current from the cold bath is given by:

\begin{align*}
\dot{Q}_i &= \hbar\omega_0 \text{Tr}\left[ \hat{J}_z \mathcal{D}_{i}(\hat{\rho}) \right]~.
\end{align*}

Inserting the expression for \(\mathcal{D}_{i}(\hat{\rho})\), using the cyclic property of the trace and introducing the notation \(\langle \hat{x} \rangle \equiv \text{Tr}[\hat{\rho} \hat{x}]\), we can rewrite this expression as:

\begin{align*}
\dot{Q}_i &= \hbar\omega_0 \gamma_i \Bigg[ (1 + n_i) \left( \langle \hat{J}_+ \hat{J}_z  \hat{J}_- \rangle - \frac{1}{2} \langle \left\{ \hat{J}_+ \hat{J}_-, \hat{J}_z \right\} \rangle \right) +\\
&\phantom{=} + n_i \left( \langle \hat{J}_- \hat{J}_z   \hat{J}_+\rangle - \frac{1}{2} \langle \left\{ \hat{J}_- \hat{J}_+, \hat{J}_z \right\} \rangle \right) \Bigg]~.
\end{align*}

Using the fact that $ \hat{J}_z$ commutes with $\hat{J}_- \hat{J}_+$ we arrive to

\begin{align*}
\dot{Q}_i &= \hbar\omega_0 \gamma_i \Big( (1 + n_i)  \langle \hat{J}_+ [\hat{J}_z,  \hat{J}_-] \rangle  + n_i  \langle \hat{J}_- [\hat{J}_z,   \hat{J}_+]\rangle \Big)~.
\end{align*}

Using the commutation relation $[\hat{J}_z, \hat{J}_\pm] = \pm \hat{J}_\pm$, we arrive to a desired final expression for the heat current \(\dot{Q}_{\rm i}\):

\begin{align*}
\dot{Q}_{i} & = \hbar\omega_0\gamma_{i} \left(- (1 + n_{i}) \langle  \hat{J}_+ \hat{J}_- \rangle +  n_{i} \langle  \hat{J}_- \hat{J}_+ \rangle \right)~,
\end{align*}

which corresponds to Eq.~\eqref{eq:heatcurrent3}.

\subsection{Analytical calculation of the heat current}
\label{ss:analitycalexpress}

\begin{figure}[th]
\centering
 \begin{overpic}[width=0.95\columnwidth]{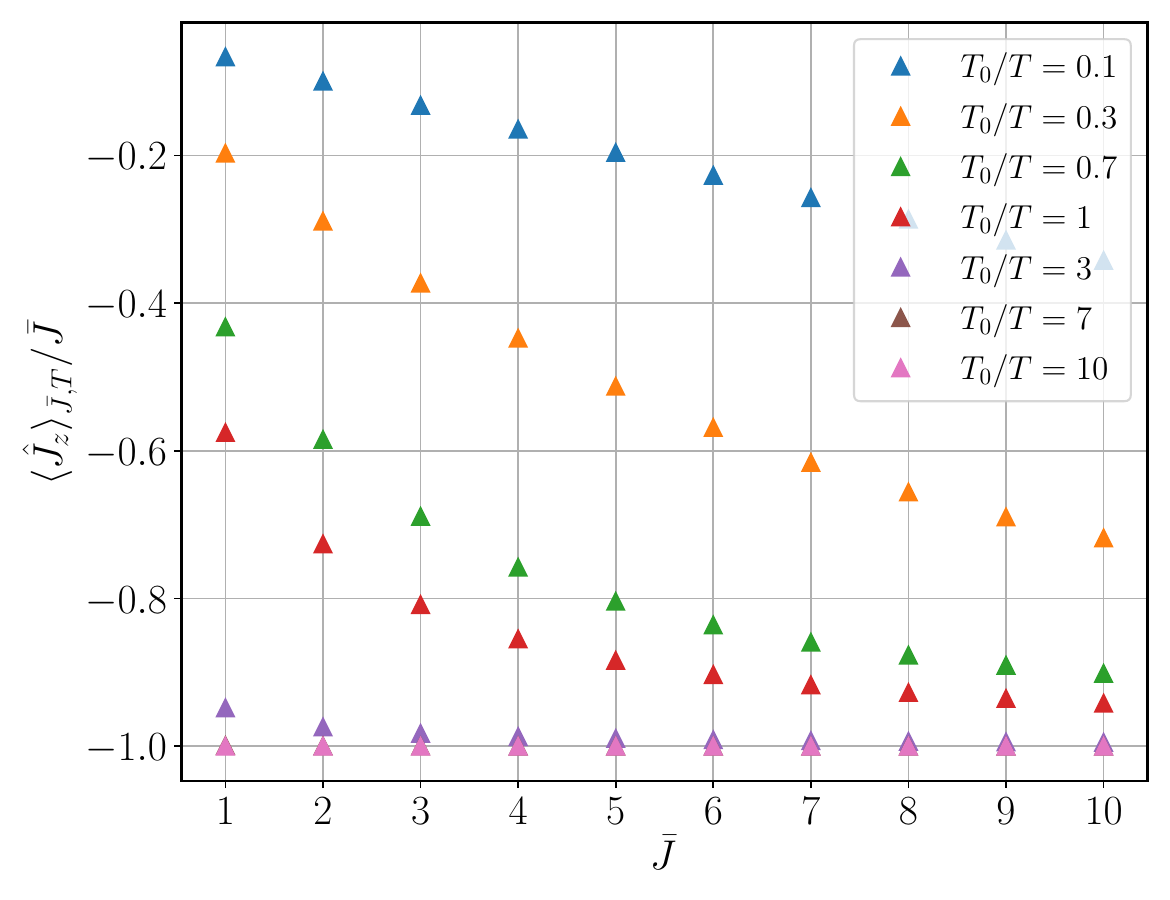}
 \end{overpic}
\caption{(Color online)This figure shows the average $\braket{\hat{J}_z}_{\bar{J},T}$ in a given subspace with $J=\bar{J}$ fixed as a function of the temperature $T$. }
\label{fig:app}
\end{figure}

In this Appendix, we detail calculations of the heat current in Eq.~\eqref{eq:heatcurrent4}. 
Essentially, one needs to calculate the average value of the collective spin inversion, $\langle -\hat{J}_z\rangle$.
Given a generic operator $\hat{x}$ its average can be computed on the steady-state density matrix $\hat{\rho}^{(\rm s)}$ in Eq.~\eqref{eq:rho} as: 
\begin{align}
\braket{\hat{x}}& =\text{Tr}\left[ \hat{\rho}^{(\rm s)} \hat{x} \right] ~,\\
&=\sum_{J} P_J \braket{\hat{x}}_{J,T}~,
\end{align}
 where $\braket{\hat{x}}_{J,T}$ denote averages within a subspace with a given $J$, over the thermal occupations $P(m_J|J)={ e^{-m\hbar\omega_0/k_{\rm B}T}}/{Z_{J,T}}$ depending on $J$ and on the temperature $T$,
\begin{align}
\braket{\hat{x}}_{J,T}& 
=\sum_{m=-J}^{J}\frac{\exp\left(-\frac{ m\hbar \omega_0}{k_{\rm B}T}\right)}{Z_{J,T}} \braket{J,m|\hat{x}|J,m}~,
\end{align}
 and ${Z_{J,T}}$ is the partition function,
\begin{align}
\label{eq:partitionF}
{Z_{J,T}}=\sum_{m=-J}^{J}  \exp\left(-\frac{ m\hbar \omega_0}{k_{\rm B}T}\right)~.
\end{align}
Then our goal reduces to calculating the average value $\langle \hat{J}_z\rangle$ in a subspace with a fixed $J$:
\begin{align}
\langle \hat{J}_z\rangle_{J,T}=\frac{\sum_{m=-J}^{J}{m \exp\left(-\frac{ m\hbar \omega_0}{k_{\rm B}T}\right)}}{Z_{J,T}}~.
\end{align}
This sum can be performed by noticing that $Z_{J,T}$ is a geometric series. The sum of this geometric series is given by:
\begin{align}
Z_{J,T} = \frac{ \exp\left(\frac{ J\hbar \omega_0}{k_{\rm B}T}\right) - \exp\left(-\frac{ (J+1)\hbar \omega_0}{k_{\rm B}T}\right)}{1 - \exp\left(-\frac{ \hbar \omega_0}{k_{\rm B}T}\right)}~.
\end{align}
Given the relation

\begin{align}
\frac{\partial}{\partial \left(\frac{1}{k_{\rm B}T}\right) }\exp\left(-\frac{ m\hbar \omega_0}{k_{\rm B}T}\right) = -\hbar\omega_0 m \exp\left(-\frac{ m\hbar \omega_0}{k_{\rm B}T}\right)~,
\end{align}
we can rewrite \( \langle \hat{J}_z \rangle_{J,T} \) as:

\begin{equation}
\label{eq:J_zDerivative}
\langle \hat{J}_z \rangle_{J,T}  = -\frac{1}{\hbar\omega_0} \frac{\partial \log(Z_{J,T})}{\partial   \left(\frac{1}{k_{\rm B}T}\right)}~.
\end{equation}
This expression provides a direct way to calculate the expectation value \( \langle \hat{J}_z \rangle_{J,T} \) by differentiating the logarithm of the partition function with respect to \( {1}/(k_{\rm B}T)  \).
By doing this derivative in Eq.~\eqref{eq:J_zDerivative} we obtain an explicit expression for \( \langle \hat{J}_z \rangle_{J,T} \):

\begin{align}
\label{eq:J_zExplicit}
\langle \hat{J}_z \rangle_{J,T}&=-J +\frac{2J+1}{1-\exp\left(\frac{ (2J+1)\hbar \omega_0}{k_{\rm B}T}\right)}+ \frac{1}{ \exp\left(\frac{ \hbar \omega_0}{k_{\rm B}T}\right)-1}~.
\end{align}
Fig.~\ref{fig:app} depicts the variation of the average $\braket{\hat{J}_z}_{J, T}$ as a function of temperature $T$. The graph illustrates a linear behavior for small effective temperatures where $J\hbar\omega_0\gg k_{\rm B} T$ and transitions to a super-linear scaling in the opposite regime, $J\hbar\omega_0\ll  k_{\rm B}T$. This behavior is indicative of the different thermal regimes influencing the collective angular momentum's $z$-component.

It is insightful to consider specific limits of Eq.~\eqref{eq:J_zExplicit}. In the regime of low-temperatures, characterized by  $\hbar\omega_0 J \gg k_{\rm B}T$, the average $\langle \hat{J}_z \rangle_{J,T}$ simplifies to:
\begin{align}
\label{eq:limit1}
\langle \hat{J}_z \rangle_{J,T}&\approx -J~.
\end{align}
This expression proves that the bound in Eq.~\eqref{eq:Q_steady_limit1} is tight in the low-temperature regime.

On the other hand, in the high-temperature regime, where $ \hbar\omega_0 J \ll k_{\rm B}T$, we obtain:
\begin{align}
\label{eq:limit2}
\langle \hat{J}_z \rangle_{J,T}&\approx-\left(\frac{\hbar\omega_0}{3k_{\rm B}T}\right)  J (1 + J) ~.
\end{align}
This equation is utilized in the main text to derive Eq.~\eqref{eq:Q_steady_limit2}.
Additionally, evaluating the ratio ${\langle \hat{J}_z \rangle_{J,T}}/({J\langle \hat{J}_z \rangle_{1/2,T}})$ in the limit $J\to \infty$ yields:
\begin{align}
\label{eq:limit3}
\frac{\langle \hat{J}_z \rangle_{J,T}}{J\langle \hat{J}_z \rangle_{1/2,T}}&\approx 2\coth\left[\frac{\hbar \omega_0}{2k_{\rm B}T}\right] ~.
\end{align}
This expression is applied in the main text to derive Eq.~\eqref{eq:Q_steady_limit1}.

\section{Calculation of $\dot{Q}^{\rm det}_{\rm h}$}
\label{appendixHeatDetuned}

In this appendix, we detail the calculation of the heat current $\dot{Q}^{\rm det}_{\rm h}$ in the detuned scenario.
In this case, each qubit interacts independently with distinct modes of the bath. The total heat current is thus given by the sum of the independent contribution of each qubit, each one determined by Eq.~\eqref{eq:heatcurrent3} setting $N=1$, with the qubit's state determined by the master equation in Eq.~\eqref{eq:DissipatorTot}. 
We assume equal strength in the couplings with both the hot and cold baths. Therefore, in the detuned scenario, the total heat current in Eq.~\eqref{eq:heatcurrent3}  can be expressed as:

\begin{equation}
\begin{split}
\label{eq:heatcurrent_detuned0}
 \dot{Q}^{\rm det}_{\rm h} &  =\sum_{k=1,2 } \hbar  \omega_0^{(k)}\gamma_{k}  \Big[- (1 + n^{(k)}_{\rm h})\langle  \hat{\sigma}^{(k)}_+ \hat{\sigma}^{(k)}_- \rangle +\\ \nonumber
 &+  n^{(k)}_{\rm h} \langle  \hat{\sigma}^{(k)}_- \hat{\sigma}^{(k)}_+ \rangle \Big]~, 
\end{split}
\end{equation}
where $\gamma_k$ represents the decay rate of the $k$-th qubit, $\hat{\sigma}^{(k)}_-$ ($ \hat{\sigma}^{(k)}_+$) are the creation (destruction) Pauli operators acting on the $k$-th qubit and the thermal occupations $n^{(k)}_{i}$ of the $i$-th bath depend on the bath temperature $T_i$ and the frequency of the $k$-th qubit:
\begin{equation}
    n^{(k)}_{i}  = \frac{1}{\displaystyle \exp\left(\frac{ \hbar \omega^{(k)}_0}{k_{\rm B}T_i}\right) - 1}~.
    \vspace{0.2cm}
\end{equation}
At steady state, the average values $\langle\hat{\sigma}^{(k)}_+ \hat{\sigma}^{(k)}_- \rangle, \langle\hat{\sigma}^{(k)}_- \hat{\sigma}^{(k)}_+ \rangle$ reduce to the thermal populations  $p^{(k)}_0,p^{(k)}_1$, as 

\begin{align}
\label{eq:populations}
\langle\hat{\sigma}^{(k)}_+ \hat{\sigma}^{(k)}_- \rangle & =p^{(k)}_1 ~,\\ 
\langle\hat{\sigma}^{(k)}_- \hat{\sigma}^{(k)}_+ \rangle & = p^{(k)}_0~, \nonumber
\end{align}
where the qubit's populations  $p^{(k)}_0,p^{(k)}_1$ are thermally occupied as dictated by Eq.~\eqref{eq:DissipatorTot}: 
$p^{(k)}_0 ={1}/\left[1+\exp\left(-\frac{ \hbar \omega^{(k)}_0}{k_{\rm B}T^*_{k}}\right)\right] $ and $
p^{(k)}_1 =1-p^{(k)}_0$.

The effective temperature $T^*_{k}$ is calculated using Eq.~\eqref{eq:DissipatorTot}, analogously to the effective temperature in the absence of the parasitic bath as given in Eq.~\eqref{eq:betastar_implicit}:

\begin{align}
\label{eq:betastar_implicit_parasite}
\frac{\gamma_k(n^{(k)}_{\rm h}+n^{(k)}_{\rm c})+\gamma_{\rm p}n^{(k)}_{\rm p}}{\gamma_k(n^{(k)}_{\rm h}+n^{(k)}_{\rm c}+2)+\gamma_{\rm p}(n^{(k)}_{\rm p}+1)} & = \exp\left(-\frac{ \hbar  \omega_0^{(k)}}{k_{\rm B}T^*_{k}}\right)~.
\end{align}
Hence, Eq.~\eqref{eq:heatcurrent_detuned0} can be recasted as:

\begin{align}
\label{eq:heatcurrent_detuned}
 \dot{Q}^{\rm det}_{\rm h} & =\sum_{k=1,2 } \hbar  \omega_0^{(k)}\gamma_{k} \left[- (1 + n^{(k)}_{\rm h})p^{(k)}_1 +  n^{(k)}_{\rm h}p^{(k)}_0 \right]~.
\end{align}

The rate for the first qubit is set at $\gamma_1=1$ GHz, as per Tab.~\ref{tab:parameters}. Detuning the second qubit from the $LC$ resonance significantly reduces its rate, based on the microscopic expression in Eq.~\eqref{eq:rates}. Hence, for the second qubit's rate $\gamma_2$, we use:
\begin{align}
\label{eq:gamma1gamma2}
\gamma_2=\frac{\gamma_1}{{1+\mathcal{Q}_i^2({\omega_0}/{\omega_{ LC}}-{\omega_{ LC}}/{\omega_0})^2}}~,
\end{align}

with $\mathcal{Q}_i$ being the quality factors of the resonators.

\bibliography{Dicke_heat_pump.bib}

\end{document}